\documentclass[12pt]{article}
\usepackage[margin=1in]{geometry}
\usepackage{amsfonts}
\usepackage{amssymb}
\usepackage{amsthm}
\usepackage{amsmath}
\usepackage{bbm}
\usepackage{array}
\usepackage{booktabs,longtable}
\usepackage[para,online,flushleft]{threeparttable}
\usepackage{threeparttablex}
\usepackage{color,tikz}
\usepackage{color}
\usepackage{dsfont}
\usepackage{enumerate}
\usepackage{epstopdf}
\usepackage{float}
\usepackage{graphicx}
\usepackage{geometry}
\usepackage[hidelinks]{hyperref}
\usepackage{indentfirst}
\usepackage{longtable}
\usepackage{lscape}
\usepackage{multirow}
\usepackage{multicol}
\usepackage{natbib}
\usepackage{rotating}
\usepackage{setspace}
\usepackage{sectsty}
\usepackage{subcaption}
\usepackage{todonotes}
\usepackage{url}
\usepackage{xcolor}
\usepackage{caption}
\usepackage{subcaption}
\usepackage{titlesec}
\usepackage{times}
\usepackage{mathptmx}
\usepackage{chngcntr}
\usepackage{amsmath,amsfonts, amssymb, dsfont, xcolor, comment}
\usepackage{mathtools}
\usepackage{natbib}
\usepackage{epsfig}
\usepackage{setspace}
\usepackage{graphicx, subcaption}
\usepackage{color}
\usepackage{mdframed}
\usepackage{float}
\usepackage[super]{nth}
\usepackage[toc,page]{appendix}
\usepackage{ragged2e}
\usepackage{makecell}
\usepackage{placeins}
\usepackage{natbib}
\usepackage{etoolbox}  
\usepackage{setspace}

\newtheoremstyle{boldprop}  
  {\topsep}   
  {\topsep}   
  {\itshape}  
  {}          
  {\bfseries} 
  {.}         
  { }         
  {\thmname{#1} \thmnumber{#2}} 
\theoremstyle{boldprop}

\newcommand{\Xomit}[1]{}

\usepackage[normalem]{ulem}
\usepackage[mathscr]{eucal}

\long\def\/*#1*/{}

\usepackage{hyperref}
\hypersetup{colorlinks,linkcolor={blue},citecolor={blue},urlcolor={blue}}

\begin{document}
\newgeometry{top=1.in, bottom=1.in, left=0.75in, right=0.75in} 
\title{AI Patents in the United States and China:\\Measurement, Organization, and Knowledge Flows\thanks{We are grateful to Lee Fleming, Kang Shi, and Xiaoyan Zhang for insightful comments and suggestions. We also appreciate the comments from the participants at seminars of PBC School of Finance at Tsinghua University, HKUST(Guangzhou), Birmingham University and the Conference on ``Frontiers in Machine Learning and Economics: Methods and Applications'' (Chicago Booth, 2025) and SFS Cavalcade Asia-Pacific Conference (2025). Zhu acknowledges the financial support from Minoru Kobayashi China Economic Research Fund, and the National Science Foundation of China (Grant No. 72273071). All remaining errors are our own.}}

\date{\today}

\author{\hspace{2mm}\\ Hanming Fang\footnote{Department of Economics, University of Pennsylvania \& the NBER, USA. Email: \href{hanming.fang@econ.upenn.edu}{hanming.fang@econ.upenn.edu}}
\hspace{5mm} Xian Gu\footnote{Department of Finance, Durham University, UK. Email: \href{xian.gu@durham.ac.uk}{xian.gu@durham.ac.uk}} \hspace{5mm} Hanyin Yan\footnote{School of Economics and Management, Tsinghua University, China. Email: \href{yanhy21@mails.tsinghua.edu.cn}{yanhy21@mails.tsinghua.edu.cn}} \hspace{5mm} Wu Zhu\footnote{School of Economics and Management, Tsinghua University, China. Email: \href{zhuwu@sem.tsinghua.edu.cn}{zhuwu@sem.tsinghua.edu.cn}} }

\maketitle 


\begin{abstract}
\normalsize
\thispagestyle{empty}
We develop a high-precision classifier to measure artificial intelligence (AI) patents by fine-tuning PatentSBERTa on manually labeled data from the USPTO’s AI Patent Dataset. Our classifier substantially improves the existing USPTO approach, achieving 97.0\% precision, 91.3\% recall, and a 94.0\% F1 score, and it generalizes well to Chinese patents based on citation and lexical validation. Applying it to granted U.S. patents (1976–2023) and Chinese patents (2010–2023), we document rapid growth in AI patenting in both countries and broad convergence in AI patenting intensity and subfield composition, even as China surpasses the United States in recent annual patent counts. The organization of AI innovation nevertheless differs sharply: U.S. AI patenting is concentrated among large private incumbents and established hubs, whereas Chinese AI patenting is more geographically diffuse and institutionally diverse, with larger roles for universities and state-owned enterprises. For listed firms, AI patents command a robust market-value premium in both countries. Cross-border citations show continued technological interdependence rather than decoupling, with Chinese AI inventors relying more heavily on U.S. frontier knowledge than vice versa.

\medskip

\noindent\textbf{JEL Classifications}: O31, O33, O34, O57, C55, G14. \\ 
\noindent\textbf{Keywords}: Artificial Intelligence; Innovation; AI Patents; Institutional Structure;  United States and China.


\end{abstract}

\restoregeometry

\newgeometry{top=1.3in, bottom=1.3in, left=1.1in, right=1.1in} 
\addtocounter{page}{-1}

\newpage
\setstretch{1.4}

\section{Introduction}
Artificial Intelligence (AI) has become a focal point of global technological competition, with the United States (US) and China emerging as the two dominant players. The past two decades have seen an exponential growth in AI patenting, with these two countries accounting for a majority of the global filings of AI patents. While the US continues to lead early innovation in the development of frontier AI research and infrastructure, China has rapidly caught up and increased its AI patent output, particularly in applied domains such as computer vision and smart manufacturing.\footnote{The emergence of the Chinese AI startup DeepSeek in December 2024 exemplifies this trend.}
These patterns suggest both convergence and divergence in the geography of AI innovation: convergence in the overall volume and intensity of R\&D, and divergence in the technological specialization and institutional foundations. Assessing the dynamics of this rivalry—whether it is characterized by convergence or divergence—is critical to understanding the future geography of innovation and productivity growth. However, empirical analysis of this competition is currently limited by a fundamental identification problem: we lack a reliable ``ruler'' to measure AI innovation accurately at scale.

To evaluate global competition in AI innovation, a critical first step is to accurately identify AI-related patents. The US Patent and Trademark Office (USPTO) released the Artificial Intelligence Patent Dataset (AIPD) in 2023 for the US \citep{giczy2022identifying, pairolero2025artificial}, which employs a Long Short-Term Memory (LSTM) model to classify AI patents. In this paper, we demonstrate that the current standard for identifying AI patents—the USPTO’s AIPD—suffers from significant measurement error in the labeled training dataset. We show that the AIPD’s classifier yields a precision of 40.5\% and a recall of 37.5\%. This implies that nearly 63\% of true AI patents are missed, and close to 60\% of the patents flagged as ``AI'' patents are true AI patents.\footnote{\textsl{Precision}, or Positive Predictive Value, summarizes the fraction of true positives (TP) among all predicted positives (TP + FP), i.e., $\textsl{Precision} = \textsl{TP}/(\textsl{TP}+\textsl{FP})$, where \textsl{FP} represents ``False Positives.'' \textsl{Recall}, or Sensitivity, or True Positive Rate, summarizes the fraction of truly positives that are identified as positives, i.e., $\textsl{Recall} = \textsl{TP}/(\textsl{TP}+\textsl{FN})$, where \textsl{FN} represents true positives that are falsely classified as negatives. \textsl{Accuracy} summarizes the fraction of cases that are correctly classified, i.e., $\textsl{Accuracy}= \textsl{(TP+TN)}/\textsl{(TP+TN+FP+FN)}.$ The \textsl{F1 Score} is the harmonic mean of \textsl{Precision} and \textsl{Accuracy,} namely, $\textsl{F1 Score}= 2\times \textsl{Precision} \times \textsl{Recall}/(\textsl{Precision} + \textsl{Recall})$.} While \citet{pairolero2025artificial} update the dataset, they do not report the performance metrics of the test set.
Such noise introduces substantial attenuation bias into any economic analysis of firm-level innovation or aggregate productivity.

Our first contribution is methodological. We  propose a new classifier of AI patents, which we call the FGYZ classifier, via applying modern natural language processing (NLP) techniques and fine-tune a suite of large language models (LLMs) to construct a novel, high-resolution dataset of AI patents. Specifically, we fine-tune \textit{PatentSBERTa}, a transformer-based language model pre-trained on patent texts, using the manually labeled AIPD seed and antiseed data from \citet{giczy2022identifying} and \citet{pairolero2025artificial}.\footnote{\textit{PatentSBERTa} was originally developed by \citet{bekamiri2021patentsberta}. Further details about the model are provided in Online Appendix ~\ref{sec:technical-detail}.} \textit{PatentSBERTa} captures rich semantic and contextual features specific to the patent domain and has been shown to outperform traditional embedding methods in classification tasks. We apply the fine-tuned classifier to the universe of patents granted by the USPTO and  China National Intellectual Property Administration (CNIPA), assigning each patent to one or more of the eight AI subfields: \textit{Machine Learning}, \textit{Natural Language Processing (NLP)}, \textit{Speech}, \textit{Vision}, \textit{Planning}, \textit{Knowledge Processing}, \textit{Hardware}, and \textit{Evolutionary Computation}.

Using our classifier, we are able to identify 876,668 unique AI patents over the period of 1976-2023 in the US and 651,630 AI patents for the period of 2010-2023 in China, matched to a total of nearly 400,000 unique individual inventors. In the AIPD seed and anti-seed data, our fine-tuned model achieves strong classification performance in seven of the eight AI subfields, with an average accuracy of 97\%, recall of 91.3\%, precision of 97.0\% and F1 score of 94.0\%. The only exception is the subfield of \textit{Evolutionary Computation}, due to limited training examples. We therefore exclude this subcategory from our subsequent empirical analyses. In general, our classifier significantly outperforms the LSTM-based model adopted by the USPTO.

We first validate our AI patent classification using citation-based connectivity measures that capture technological relatedness across patent groups. Specifically, we partition patents based on the joint outcomes of two classification systems: the USPTO’s Artificial Intelligence Patent Dataset (AIPD) and our classifier (FGYZ). Patents classified as AI by \textit{both} classifiers form a high-confidence positive set. We show that patents identified as AI only by FGYZ show systematically stronger citation connectivity to this benchmark set, and significantly weaker connectivity to patents classified as non-AI by both classifiers. This pattern holds across AI subfields and is particularly pronounced in core domains such as machine learning, natural language processing, and planning. In contrast, patents classified as AI only by the USPTO  show weaker links to the AI benchmark and stronger connections to non-AI technologies. These results indicate that patents captured by our classifier are more tightly embedded in the AI knowledge network.

We complement citation-based validation with a lexical similarity analysis that compares the technical language used across patent groups. Using TF–IDF–weighted word distributions, we measure how closely the vocabulary of each group aligns with that of the high-confidence AI benchmark, i.e., patents jointly identified as AI by both the USPTO and the FGYZ classifiers. Across nearly all AI subfields, patents identified as AI only by FGYZ show substantially higher lexical similarity to the benchmark than those identified as AI only by the USPTO classification. The differences are especially pronounced in machine learning, planning, and hardware-related AI. Together, citation and lexical evidence show that our classifier identifies patents that are not only more strongly connected to established AI technologies through citations but also more closely aligned in technical language with high-confidence AI patents jointly identified as AI by both classification systems.

We further examine the external validity of the FGYZ classifier by extending the analysis to Chinese patents. Using cross-country citation connectivity and lexical similarity measures, we compare Chinese patents classified by our classifier as AI with US patents classified as AI and non-AI. We find that Chinese patents identified as AI exhibit substantially stronger citation connectivity to US AI patents and weaker connectivity to US non-AI patents across all AI subfields. These patterns are particularly pronounced in core AI domains such as machine learning, natural language processing, and speech. Consistent with the citation evidence, Chinese AI patents also display significantly higher lexical similarity to US AI patents than Chinese non-AI patents, suggesting close alignment in technical vocabulary with the global AI knowledge frontier. These results together show that the classifier trained on US patents generalizes well to Chinese data and successfully identifies Chinese AI patents that are technologically connected to and semantically aligned with established AI innovation in the US.

Using the FGYZ classification, we document a rapid expansion of AI patenting activity in both the US and China over the past decade. The number and share of AI patents increase significantly in both countries, with a particularly strong acceleration after the mid-2010s. While the US leads early in AI patenting, China overtakes the US in the total annual number of AI patents in recent years, reflecting broader growth in Chinese inventive activity. Despite differences in scale, the distribution of AI patents across technology subfields is broadly similar in the two countries, with planning, vision, and hardware accounting for the largest shares. At the same time, divergences arise in specific AI subfield, particularly natural language processing, where US patenting expands earlier, while Chinese patenting accelerates sharply after 2020.

The geographic and institutional landscape of AI innovation also differs systematically between the two countries. In both the US and China, AI patenting activity is highly clustered in major innovation hubs and remains persistently concentrated over time, even as secondary clusters gradually emerge. In the US, AI innovation is dominated by a small set of large multinational technology firms—such as IBM, Microsoft, Google, and Amazon—across nearly all AI subfields, indicating a broad technological scope and stable incumbent leadership. In contrast, China shows a more heterogeneous institutional landscape: Along with leading private technology firms such as Tencent, Baidu, and Huawei, state-owned enterprises and universities play a prominent role in AI patenting, particularly in hardware- and application-oriented subfields. 

Despite these similarities in initial clustering, the spatial diffusion of AI innovation differs between the two countries. In the US, AI patenting remains tightly anchored in early hubs, with limited geographic expansion over time, consistent with a mature innovation landscape. In contrast, China experiences rapid spatial diffusion, as AI activity spreads quickly from pioneering cities to a growing set of secondary locations. This pattern reflects faster catch-up outside initial hubs and a widening geographic footprint of AI innovation. These dynamics suggest convergence in the technological composition of AI innovation across the US and China, alongside divergence in its geographic diffusion and institutional organization.

Next, we evaluate the AI patent value using stock market reactions to patent disclosures following \cite{kogan2017technological}, and we find that AI patents are associated with systematically higher economic value than non-AI patents in both the US and China. This AI-related valuation premium is present across all subfields of AI technology in both countries. Although the overall level of patent value is higher in the US than in China, reflecting differences in market size and capitalization, the relative premium associated with AI patents is robust in both settings. The magnitude of the AI premium varies across technology subfields and between countries, with larger premia concentrated in data- and software-intensive domains such as machine learning and natural language processing, and more modest premia in more hardware-oriented AI technologies.

For patents granted to non-public entities, their economic value cannot be directly inferred from stock market reactions. This raises an important question: do non-market institutions—such as universities and state-owned enterprises (SOEs)—meaningfully contribute to technological progress? This issue has sparked growing debate among scholars and policymakers.\footnote{For a recent discussion, see ``Study: Industry Now Dominates AI Research,'' MIT Sloan School of Management, \href{https://mitsloan.mit.edu/ideas-made-to-matter/study-industry-now-dominates-ai-research}{https://mitsloan.mit.edu/ideas-made-to-matter/study-industry-now-dominates-ai-research}.}

We discover substantial cross-country differences in the structure of knowledge flows. In the US, academic institutions largely operate as intellectual enclaves: university-generated patents are predominantly cited by other academic entities, with limited direct engagement from private firms—reflecting a classic ``ivory tower'' dynamic. In contrast, Chinese research institutions and SOEs maintain dense reciprocal linkages with industry. Private firms in China are more likely to cite patents from non-market institutions than from other private firms, and academic and corporate sectors frequently cite each other. These findings challenge the view that non-market innovation in China is primarily administrative or low-value and instead highlight the strategic role of state-sector actors in generating economically relevant AI technologies.

Finally, despite rising geopolitical tensions, we find little evidence of technological decoupling in AI. Instead, cross-border citation patterns point to sustained, and in some dimensions, intensifying, technological coupling between the US and China. Chinese AI inventors are highly influenced by recent US frontier technologies, suggesting a continued reliance on global knowledge leaders. Knowledge flows in the opposite direction are more selective, and US inventors cite Chinese patents more intensively outside core AI domains. Together, these patterns suggest that AI competition between the two countries is not characterized by isolation, but by asymmetric cross-border learning embedded within an increasingly competitive global innovation environment.

The remainder of the paper is structured as follows. Section \ref{sec:ai-prediction} describes the patent datasets and introduces the large language model–based methodology used to classify AI patents. Section \ref{sec:comparison} validates the classification using citation-based connectivity and lexical similarity analyses. Section \ref{sec:extension-china} extends the classification to Chinese patents and provides validation out-of-sample. Section \ref{sec:facts} presents a comparative analysis of AI innovation in the US and China, examining time trends, geographic distribution, institutional ownership, and the economic value of AI relative to non-AI patents. Section \ref{sec:conclusion} concludes.

\section{Using LLM To Classify AI Patents}\label{sec:ai-prediction}
\subsection{Patent Datasets}
Our primary data set comprises patent records obtained from the USPTO and CNIPA. We focus on \textit{utility} patents granted by the USPTO and \textit{invention} patents granted by the CNIPA, as these categories represent the most substantive and economically significant forms of patents in each jurisdiction \citep{fang2021patentquality}. The data set includes approximately 7.7 million USPTO patents granted between 1976 and 2023, and 5.4 million CNIPA patents granted between 2010 and 2023. We restrict our sample to patents granted after 1976 for two primary reasons. First, the emergence of AI-related innovation is a relatively recent phenomenon, and patents filed before 1976 are unlikely to be relevant to the technological frontier in AI. Second, full-text data, including abstracts and claims, which are essential inputs for our classification algorithm, are readily available for patents granted after 1976, allowing us to conduct a consistent document-level analysis. Although there are procedural differences between the USPTO and CNIPA examination systems, previous research indicates that they are broadly comparable in terms of substantive examination standards and practices \citep{Han2024}. This comparability supports the validity of cross-country analyses based on our combined dataset.

Each patent record in our data set includes detailed information on the application date, publication date, grant date, inventor(s), assignee (s), international patent classification (IPC) codes, and textual contents, including abstracts and claims. This rich information allows for a comprehensive analysis of innovation activity. We retrieve both forward and backward citations for the patents in our dataset from Google Patents. We classify patent assignees in China into three types: institutions, privately owned enterprises (POEs), or state-owned enterprises (SOEs). The patent assignees in the US are classified into two types: institutions and enterprises (almost all are POEs). We retrieve this information from China's State Administration for Industry and Commerce (SAIC), following \cite{allen2024centralization}. 

\paragraph{AI Patents.} Neither the USPTO nor the CNIPA provides an official classification for patents related to AI. To address this gap, USPTO researchers developed the Artificial Intelligence Patent Dataset (AIPD), released in 2023, which covers approximately 15.4 million US patents and published patent applications (PGPubs) filed between 1976 and 2023 \citep{pairolero2025artificial}.

The construction of the AIPD began with the manual annotation of a subset of patents, which were classified into eight core AI domains: \textit{Machine Learning}, \textit{Evolutionary Computation}, \textit{Natural Language Processing}, \textit{Speech}, \textit{Vision}, \textit{Knowledge Processing}, \textit{Planning}, and \textit{Hardware} \citep{giczy2022identifying,pairolero2025artificial}. Appendix \ref{tab:AIpatent_definition_example} presents the definition and an example for each subcategory.  The data set was developed through a rigorous multi-stage process: “precise core‐sample definition $\rightarrow$ systematic scope expansion $\rightarrow$ strict negative‐case screening $\rightarrow$ manual annotation and validation”. The AIPD data are a labeled training set consisting of a seed set of positive examples (AI patents) and an anti-seed set of negative examples (non-AI patents). 

Using this manually labeled data set, \citet{giczy2022identifying} and \citet{pairolero2025artificial} trained a Long Short-Term Memory (LSTM) model to classify AI patents. However, the predictive performance of the model is limited. Although the reported test-set \textsl{accuracy} is approximately 87\%, this figure is misleading due to the overwhelming dominance of non-AI patents in the sample.\footnote{For example, if more than 80\% of patents are non-AI, a naive classifier that always predicts “non-AI” would already achieve an accuracy rate of 80\%.} A closer look at class-specific metrics reveals more serious deficiencies: the \textsl{precision} of the model is only 40.5\%, which implies that less than half (4.05 out of 10) of the patents it classifies as AI are truly related to AI. The \textsl{recall rate} is similarly low at 37.5\%, which means that the model fails to identify more than 60\% of the actual AI patents. Consequently, \textsl{F1 score}, that is, the harmonic mean of precision and recall, is just 39\%, raising significant concerns about the reliability of the model, particularly in applications focused on AI innovation at the firm level and its economic implications.

\subsection{FGYZ Classifier and AI Classifications}

To improve the precision of classification, we build on the manually labeled dataset employing a transformer-based architecture. Specifically, we construct our model, which we will refer to as the FGYZ classifier  hereafter, by fine-tuning \textit{PatentSBERTa} \citep{bekamiri2021patentsberta,BiermannHuber2024}, a domain-adapted variant of \textit{Sentence-BERT} trained in large-scale patent corpora, including abstracts, claims, and full text descriptions. The model is optimized using a contrastive learning objective designed to capture contextual and semantic relationships specific to the patent domain. Previous work \citep{bekamiri2021patentsberta} shows that \textit{PatentSBERTa} outperforms conventional embedding methods in classification, retrieval, and clustering tasks. The technical details of the fine-tuning procedure are provided in the Online Appendix~\ref{sec:technical-detail}. 

We apply \textit{PatentAIBERTa} to the original AIPD seed and anti-seed datasets developed by \citet{giczy2022identifying} and \citet{pairolero2025artificial}, and train a binary classifier for AI patents and their subcategories. 
For each AI subcategory, we partition the manually labeled seed (positive) and anti-seed (negative) sample into an 80/20 training-test split. Within the training set, we implement five-fold cross-validation to tune the hyperparameters and determine early stopping criteria (see Appendix~\ref{sec:technical-detail} for details).

\begin{center}
[Insert Table~\ref{tab:train_and_test_accuracy} About Here]
\end{center}

Panel A in Table \ref{tab:train_and_test_accuracy} summarizes the training and test data sets used to fine-tune our AI patent classifier, and reports the number of training and test samples, together with the counts of positive and negative cases for each AI subcategory. Panel B presents the classification performance of the test set, including accuracy, precision, recall, and F1 score by category. 

\begin{figure}[htbp]
\centering
\caption{Model Performance by AI subcategories}
\label{fig:model_performance_visual}
\vspace{5mm}
\caption*{This figure presents the distribution of predicted probabilities conditional on the true label of patents in the testing set using US patents. For each AI subcategory, we report boxplots of the conditional distributions \(P(\hat{y} = 1 \mid y = 1)\) and \(P(\hat{y} = 1 \mid y = 0)\), where \(y = 1\) denotes a true positive and \(\hat{y} = 1\) indicates that the classifier predicts the patent as belonging to that AI subcategory. In the following subplots, ``Yes" indicates \(y = 1\) while ``Not" corresponds to \(y = 0\). In all subcategories except \textit{Evolutionary Computation}, the distributions of \(P(\hat{y} = 1 \mid y = 1)\) and \(P(\hat{y} = 1 \mid y = 0)\) are tightly concentrated near 1 and 0, respectively, suggesting strong classification performance and clear separability between AI and non-AI patents for each subcategory. However, in the case of \textit{Evolutionary Computation}, the distribution of \(P(\hat{y} = 1 \mid y = 0)\) is centered around 0.45 with high variance, reflecting poor distinction between positive and negative cases.}
\vspace{5mm}
\makebox[\textwidth][c]{
\includegraphics[width=1.1\textwidth]{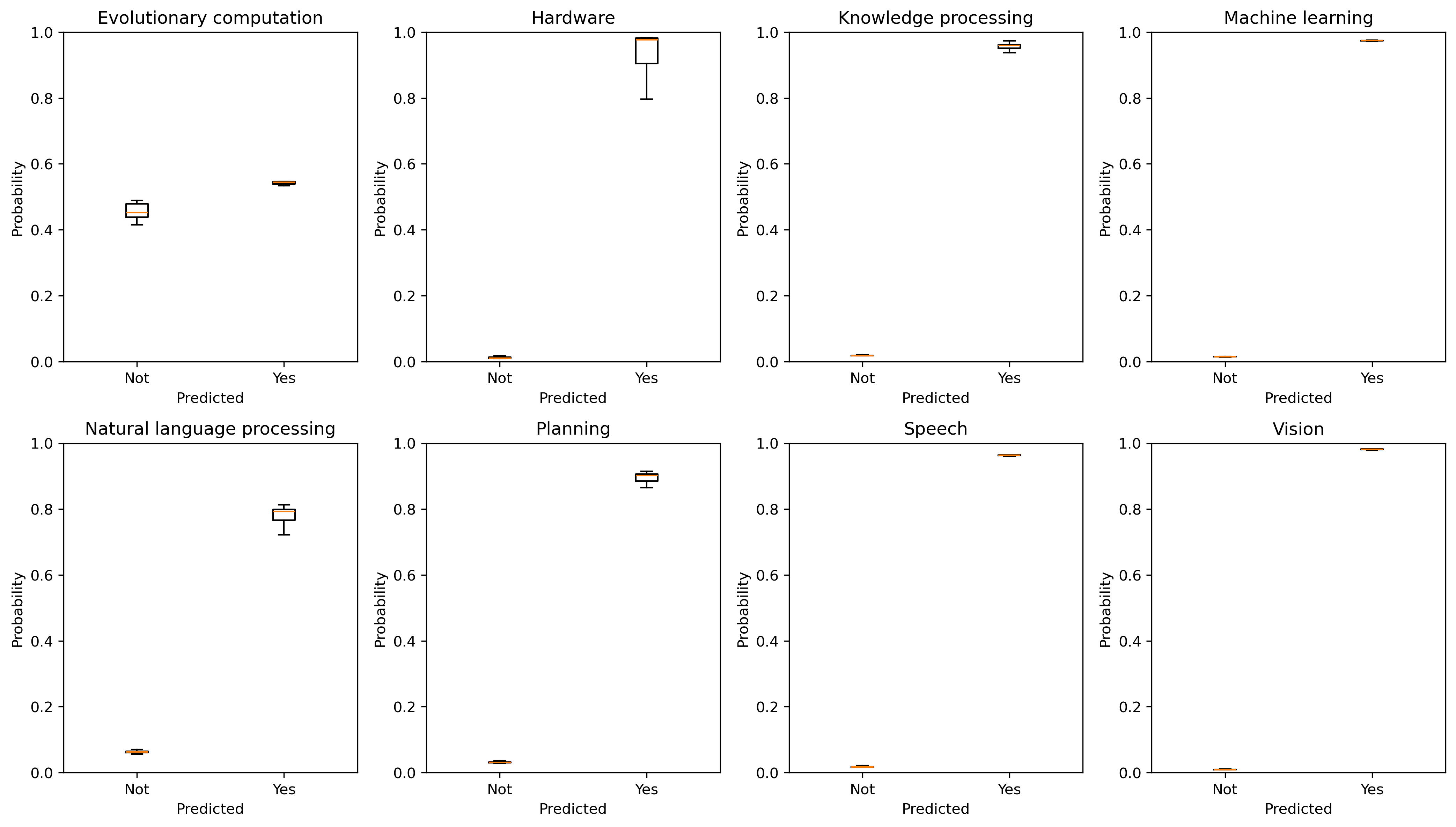}
}
\end{figure}

To compare the performance of our model with the USPTO's LSTM-based approach, Panel C provides aggregate classification metrics across all subcategories. The results show that, relative to the USPTO's model, our classifier achieves substantially higher precision, recall, and F1 score. Specifically, our model reaches an overall precision of 97.0\%, recall of 91.3\%, and F1 score of 94.0\%, significantly improving the USPTO reported values of 40.5\%, 37.5\% and 39\%, respectively. This highlights the superior reliability of our method in identifying AI patents, particularly in imbalanced classification settings. Importantly, classification performance is consistently high in seven of the eight subcategories. In each case, our model significantly outperforms the accuracy reported by the USPTO \citep{giczy2022identifying,pairolero2025artificial}. The only exception is the \textit{Evolutionary Computation} category, where performance deteriorates due to the limited size of the training sample, comprising only 128 labeled observations (25 positives and 103 negatives). This has resulted in the model's lower predictive accuracy in this subcategory. 
\begin{center}
[Insert Figure \ref{fig:model_performance_visual} About Here]
\end{center}
We further validate the discriminative precision of our model by examining the distribution of predicted probabilities conditional on the ground truth labels. If the LLM has effectively internalized the semantic boundaries of AI technologies, we would expect the conditional predicted probabilities to show sharp polarization: the probability assigned to true AI patents, $P(\hat{y} = 1 \mid y = 1)$, should converge towards 1, while the probability assigned to non-AI patents, $P(\hat{y} = 1 \mid y = 0)$, should converge towards 0.

Figure \ref{fig:model_performance_visual} plots these conditional distributions for each subcategory. The empirical results strongly corroborate our expectations. For the vast majority of subfields, the classifier demonstrates a decisive separation, with the probability mass tightly concentrated at the extremes of the unit interval. This confirms that the model detects distinct semantic signals. The only exception is \textit{Evolutionary Computation}, where the model shows substantial uncertainty, with predictions for non-AI patents loosely centered around 0.45. Consequently, to ensure that our stylized facts are not driven by classification noise, we exclude this subcategory from subsequent economic analysis.

Figure~\ref{fig:subcategory_percentage} illustrates the distribution of AI patents by the number of AI subfields to which they belong. To construct this measure, we train eight classifiers corresponding to distinct AI subcategories and apply them to patent texts to determine whether a patent belongs to each subfield. Because the classification is multi-label, patents can be assigned to multiple AI subfields simultaneously. The figure therefore reports the percentage of AI patents that fall into one, two, three, or more subfields. We exclude the evolutionary computation category (\textit{evo}) due to its small sample size, so the number of subfields ranges from one to seven. 

The figure compares three datasets. ``US-AIPD'' denotes classifications provided by the USPTO AI Patent Dataset, while ``US-FGYZ'' and ``China-FGYZ'' refer to classifications generated by our FGYZ classifier applied to the US and Chinese patents, respectively. The results indicate that AI innovation frequently spans multiple technological domains: 46\% of AI patents in the US and 49\% in China are classified into at least two subcategories. This pattern highlights the interdisciplinary nature of AI technologies.

\section{Comparison of Classifiers}\label{sec:comparison}
To provide a more intuitive understanding of the model's performance relative to the benchmark, we examine the cross-system consistency between the USPTO AIPD and our FGYZ classifier. We partition the patent corpus into four mutually exclusive groups based on the joint classification outcomes: (1) \textit{Dual-Positive}, identified as AI-related patents by both classifiers; (2) \textit{AIPD-Positive}, identified only by the AIPD classifier; (3) \textit{FGYZ-Positive}, identified only by our FGYZ classifier; and (4) \textit{Dual-Negative}, rejected by both.

The implicit motivation for this decomposition is that the \textit{Dual-Positive} intersection represents the highest-confidence ``consensus'' set of true AI patents. This allows us to formulate a clear intuitive test: if our classifier is successfully identifying genuine AI patents that the AIPD misses, we would expect the \textit{FGYZ-Positive} group to be technologically similar to the \textit{Dual-Positive} consensus and dissimilar to the \textit{Dual-Negative} baseline. Conversely, if the benchmark suffers from false positives, the \textit{AIPD-Positive} group should exhibit a weaker resemblance to the consensus set. This framework allows us to characterize the nature of the divergence between the two measures. All comparisons are conducted within the seven AI technology categories to ensure comparability across subfields, excluding \textit{Evolutionary Computation}.

\subsection{Citation-Based Connectivity Analysis}
\label{sec:connectivity}

To validate the semantic accuracy of our classifier, we turn to the principle of ``\textit{technological homophily}:'' genuine AI innovations should disproportionately build upon the existing AI frontier. We examine this by constructing a size-adjusted measure of citational connectivity. 

The core challenge in measuring connectivity is the vast size imbalance between the AI and non-AI patent universes. A raw citation count would be dominated by noise from the non-AI sector. To correct for this, we define the \textit{Directional Citation Preference} from group $A$ to group $B$ as the observed citation probability normalized by the random baseline:

\begin{equation}
R\left(A\rightarrow B\right)=\frac{\#\ \text{of}\ A' \text{s\ citation\ to\ patents\ in}\ B\ /\ \#\ \text{of}\ A' \text{s\ citation\ to\ patents\ outside}\ A}{\#\ \text{of\ patents\ in}\ B\ /\ \#\ \text{of\ patents\ outside}\ A}.
\label{eq:R_A_to_B}
\end{equation}

This measure captures how strongly patents in the group $A$ are connected to patents in the group $B$ through citations. Specifically, it compares the fraction of citations made by patents in $A$ that point to patents in $B$ with the share of patents that belong to the group $B$ in the population outside $A$.
A value $R(A \rightarrow B) > 1$ means that patents in $A$ cite patents in $B$ more frequently than would be expected if citations were randomly assigned proportional to the size of the group. We define the reverse-direction preference $R(B \rightarrow A)$ in the same way.

To obtain a symmetric measure of cross-group connectivity, we average the two directional preferences:

\begin{equation}
C(A,B) = 
\frac{
R(A \rightarrow B) + R(B \rightarrow A)
}{2}.
\label{eq:C_AB}
\end{equation}

The statistic $C(A,B)$ captures the mutual citation connectivity between the groups $A$ and $B$. Higher values indicate a stronger technological connectedness.

We compute citation-based connectivity measures for all pairwise combinations of the four patent groups defined by the joint AIPD--FGYZ classification system: \textit{AIPD-Positive}, \textit{FGYZ-Positive}, \textit{Dual-Positive}, and \textit{Dual-Negative}. For each pair of groups, we construct the symmetric connectivity measure $C(A,B)$ defined in Equation~(\ref{eq:C_AB}). This approach quantifies how frequently patents in one group cite patents in another, relative to what would be expected under random citation proportional to group size. Comparing these connectivity patterns allows us to evaluate how closely the \textit{AIPD-Positive} and \textit{FGYZ-Positive} patents are linked to the high-confidence \textit{Dual-Positive} set and how distinct they are from the \textit{Dual-Negative} group.

\begin{figure}[htbp]
\centering
\caption{Citation-Based Connectivity of AIPD and FGYZ Patents to Dual-Positive (Negative) Patents}
\label{fig:us_mutual_cite}
\vspace{2mm}
\caption*{
This figure shows the citation-based connectivity of \textit{AIPD-Positive} and \textit{FGYZ-Positive} patents to dual-positive or dual-negative group (benchmark group). Panel (a) reports the connectivity of each group to the \textit{Dual-Positive} set, defined as patents jointly classified as AI by both systems and therefore treated as high-confidence AI patents. Panel (b) reports connectivity to the \textit{Dual-Negative} group. Connectivity is computed using the bidirectional citation preference measure defined in Section~\ref{sec:connectivity}, and reflects how intensively patents in each group cite patents in the benchmark group relative to random citation proportional to group size. Higher values indicate stronger technological relatedness. Across AI subfields, \textit{FGYZ-Positive} patents exhibit systematically stronger connectivity to \textit{Dual-Positive }patents and weaker connectivity to \textit{Dual-Negative} patents than \textit{AIPD-Positive }patents, providing evidence of higher classification confidence.
}
\vspace{2mm}
\begin{subfigure}[t]{\textwidth}
    \centering
    \caption{To Dual-Positive}
    \includegraphics[width=0.8\linewidth]{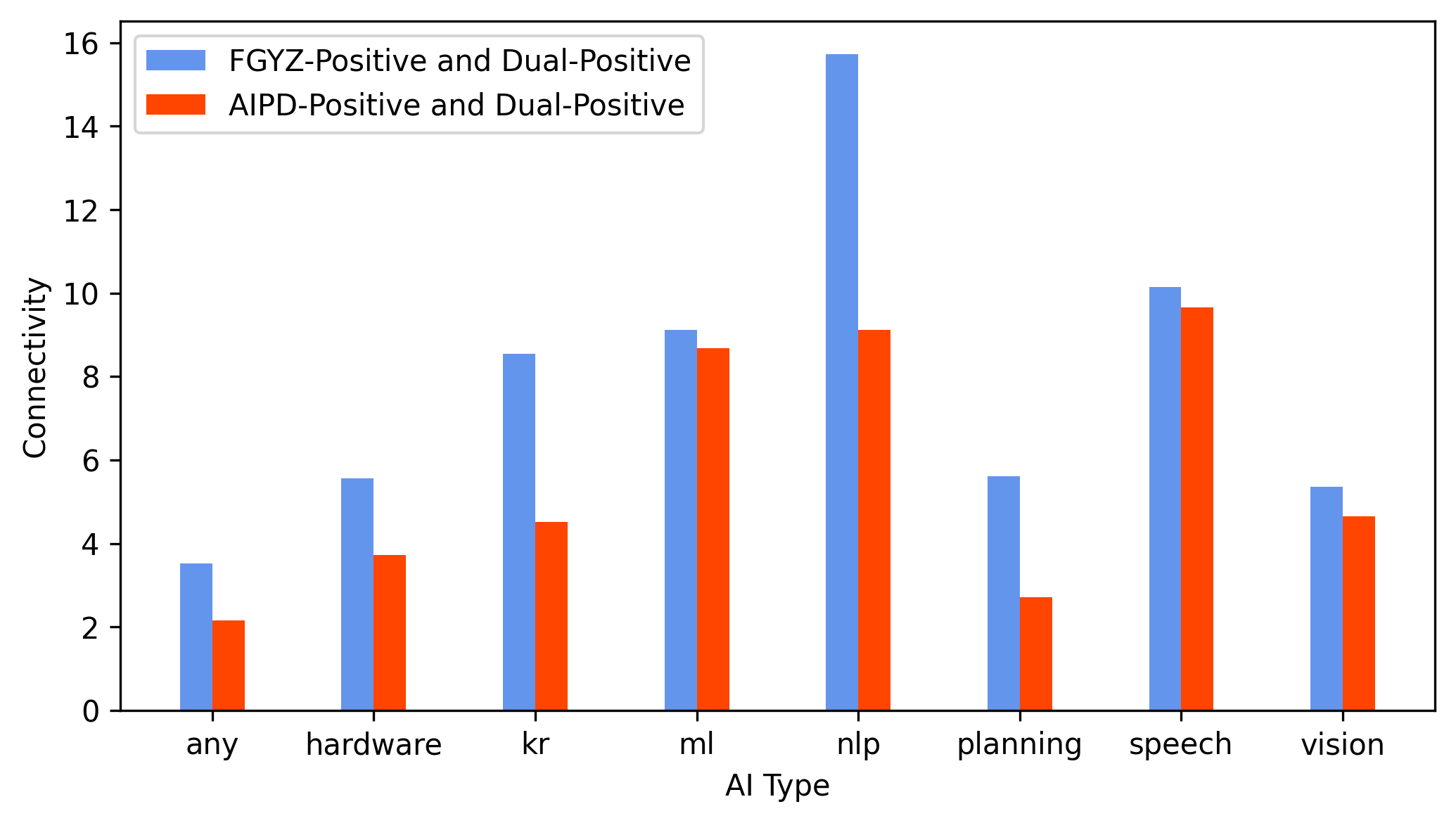}
\end{subfigure}

\vspace{2mm}

\begin{subfigure}[t]{\textwidth}
    \centering
    \caption{To Dual-Negative}
    \includegraphics[width=0.8\linewidth]{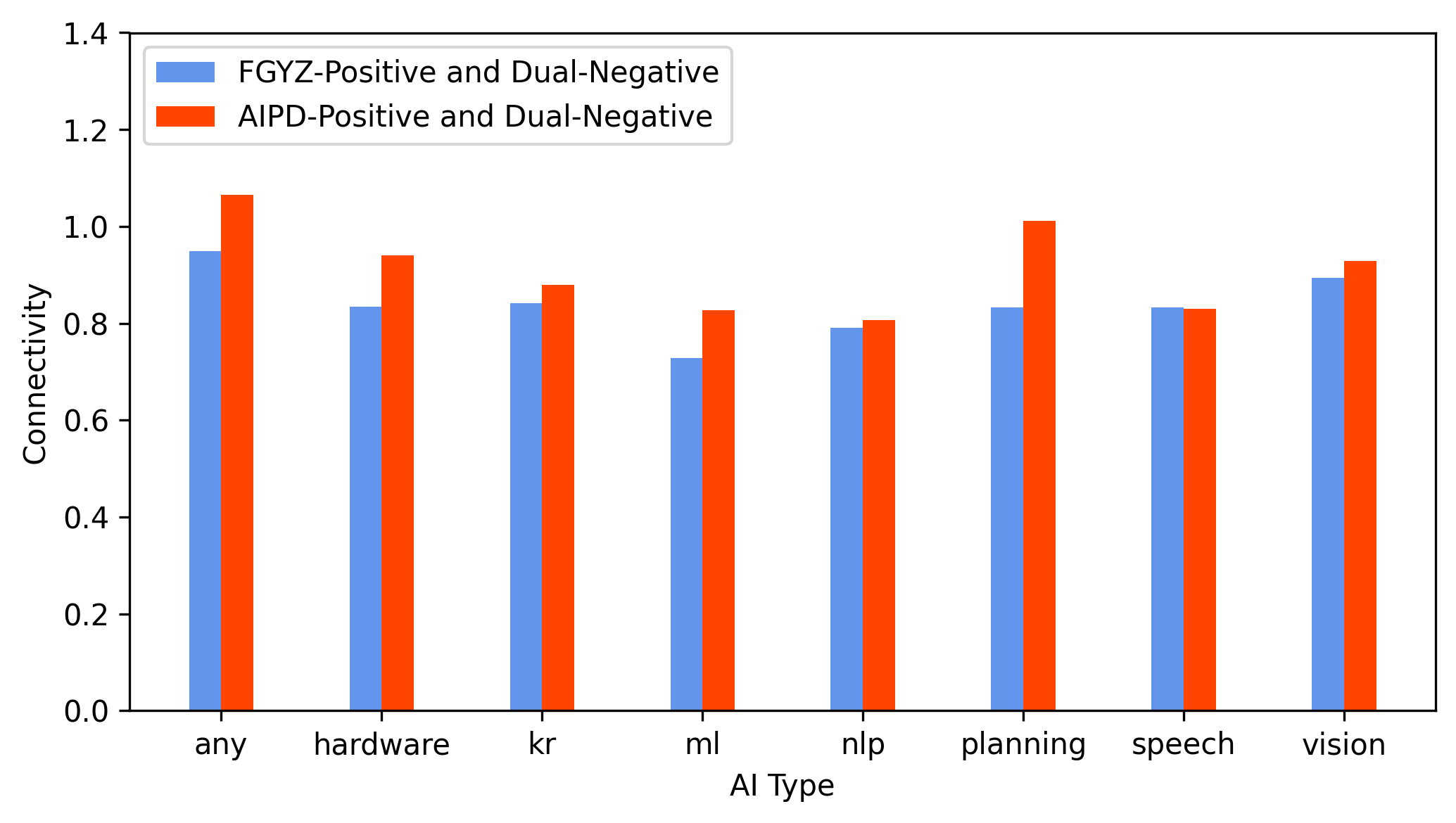}
\end{subfigure}
\end{figure}

Figure~\ref{fig:us_mutual_cite} reports citation-based connectivity between \textit{AIPD-Positive} and \textit{FGYZ-Positive} patents and the two benchmark groups. Panel (a) shows connectivity to the \textit{Dual-Positive} set, which represents high-confidence AI patents jointly identified by both systems. Across all AI subfields, \textit{FGYZ-Positive} patents exhibit substantially stronger connectivity to the \textit{Dual-Positive} benchmark than \textit{AIPD-Positive} patents. This pattern is especially pronounced in core AI domains such as Natural Language Processing (NLP), Knowledge Processing (KR), and Planning, where the connectivity of \textit{FGYZ-Positive} patents is substantially higher. These results indicate that patents identified only by FGYZ are more closely linked, through citations, to the consensus set of AI patents in the \textit{Dual-Positive} group.

Panel (b) reports connectivity to the \textit{Dual-Negative} group, which is unlikely to contain AI-related patents. In contrast to Panel (a), \textit{FGYZ-Positive} patents display systematically weaker connectivity to \textit{Dual-Negative} patents than \textit{AIPD-Positive} patents across most AI subfields. Connectivity values for \textit{FGYZ-Positive} patents are consistently below one, indicating that these patents cite non-AI technologies less frequently than would be expected under random citation.

\begin{center}
    [Insert Figure~\ref{fig:us_mutual_cite} About Here]
\end{center}

\subsection{Lexical Distribution Similarity Analysis}
\label{sec:lexical_similarity}

While citation connectivity captures the \textit{economic} structure of knowledge diffusion, it provides less information on \textit{semantic} consistency; a patent may cite an AI algorithm for a purely peripheral application. We introduce a complementary analysis of \textit{lexical} similarity that sheds new insights into our classifier. One thing worth mentioning is that raw textual comparison is often fraught with measurement error due to ubiquitous legal boilerplate and generic technical vocabulary. To mitigate this, we construct a discriminatory weighting scheme that explicitly penalizes terms common in the non-AI universe, allowing us to isolate the distinctive vocabulary of the technological frontier.

We first construct a ``background noise'' distribution from the \textit{Dual-Negative} group (patents rejected by both systems) which is unlikely to contain AI patents. For each word $w$ in the vocabulary, we calculate a penalty weight using an inverse-document-frequency (IDF) formulation \footnote{The IDF weighting scheme is a standard methodological tool in textual analysis for filtering out generic syntax (or ``boilerplate'') while amplifying semantically distinct terms. This approach has been widely adopted in the economics of innovation and finance literature to capture technological novelty and sentiment; see, for example, \cite{kelly2021measuring} and \cite{cong2024automation}.}:

\begin{equation}
\mathrm{IDF}_{\text{DN}}(w)
=
\log
\left(
\frac{\sum_{w'} \mathrm{TF}_{\text{DN}}(w')}{\mathrm{TF}_{\text{DN}}(w) + 1}
\right).
\label{eq:IDF_DN}
\end{equation}
where $\mathrm{TF}_{\text{DN}}(w)$ denotes the total frequency of terms for word $w$ in the \textit{Dual-Negative} group.This weighting scheme down-weights words that appear very frequently in the \textit{Dual-Negative} group and up-weights words that are relatively rare in that background, thus emphasizing terms that are more distinctive of AI-related content.

For each focal patent group $A$ in \textit{AIPD-Positive}, \textit{FGYZ-Positive}, and \textit{Dual-Positive}, we then construct a TF-IDF–reweighted word distribution as
\begin{equation}
\widetilde{P}_A(w)
=
\frac{\mathrm{TF}_A(w)}{\sum_{w'} \mathrm{TF}_A(w')}
\cdot
\mathrm{IDF}_{\text{DN}}(w),
\label{eq:P_tilde_A}
\end{equation}
where $\mathrm{TF}_A(w)$ denotes the total frequency of terms for word $w$ in group $A$.

To quantify the lexical similarity between two groups $A$ and $B$, we compute the inner product of their TF-IDF–reweighted distributions:
\begin{equation}
S(A,B)
=
\sum_{w} \widetilde{P}_A(w)\,\widetilde{P}_B(w).
\label{eq:similarity_inner_product}
\end{equation}
The statistic $S(A,B)$ increases in the overlap of high-weight words between groups $A$ and $B$, thus serving as a measure of lexical similarity. In our application, we use this measure to compare how similar the \textit{AIPD-Positive} and \textit{FGYZ-Positive} groups are to the high-confidence \textit{Dual-Positive} group in terms of their vocabulary profiles.

\begin{figure}[htbp]
\centering
\caption{Lexical Similarity of AIPD and FGYZ Patents to Dual-Positive (Negative) Patents}
\label{fig:us_lexical_similarity}
\vspace{2mm}
\caption*{
This figure shows the lexical similarity of \textit{AIPD-Positive }and \textit{FGYZ-Positive} patents to the dual-positive or dual negative (benchmark). Panel (a) reports the similarity of each group to the \textit{Dual-Positive} set, defined as patents jointly classified as AI by both systems and therefore treated as high-confidence AI patents. Panel (b) reports similarity to the \textit{Dual-Negative} group. Lexical similarity is measured as the inner product between the TF--IDF--weighted word distributions between each group and the benchmark patents, following the procedure described in Section~\ref{sec:lexical_similarity}. Higher values indicate closer alignment in vocabulary usage. Across AI subfields, \textit{FGYZ-Positive }patents consistently show higher lexical similarity to the \textit{Dual-Positive} group than \textit{AIPD-Positive} patents, suggesting stronger alignment with the high-confidence AI benchmark.
}
\vspace{2mm}
\begin{subfigure}[t]{\textwidth}
    \centering
    \caption{To Dual-Positive}
    \includegraphics[width=0.8\linewidth]{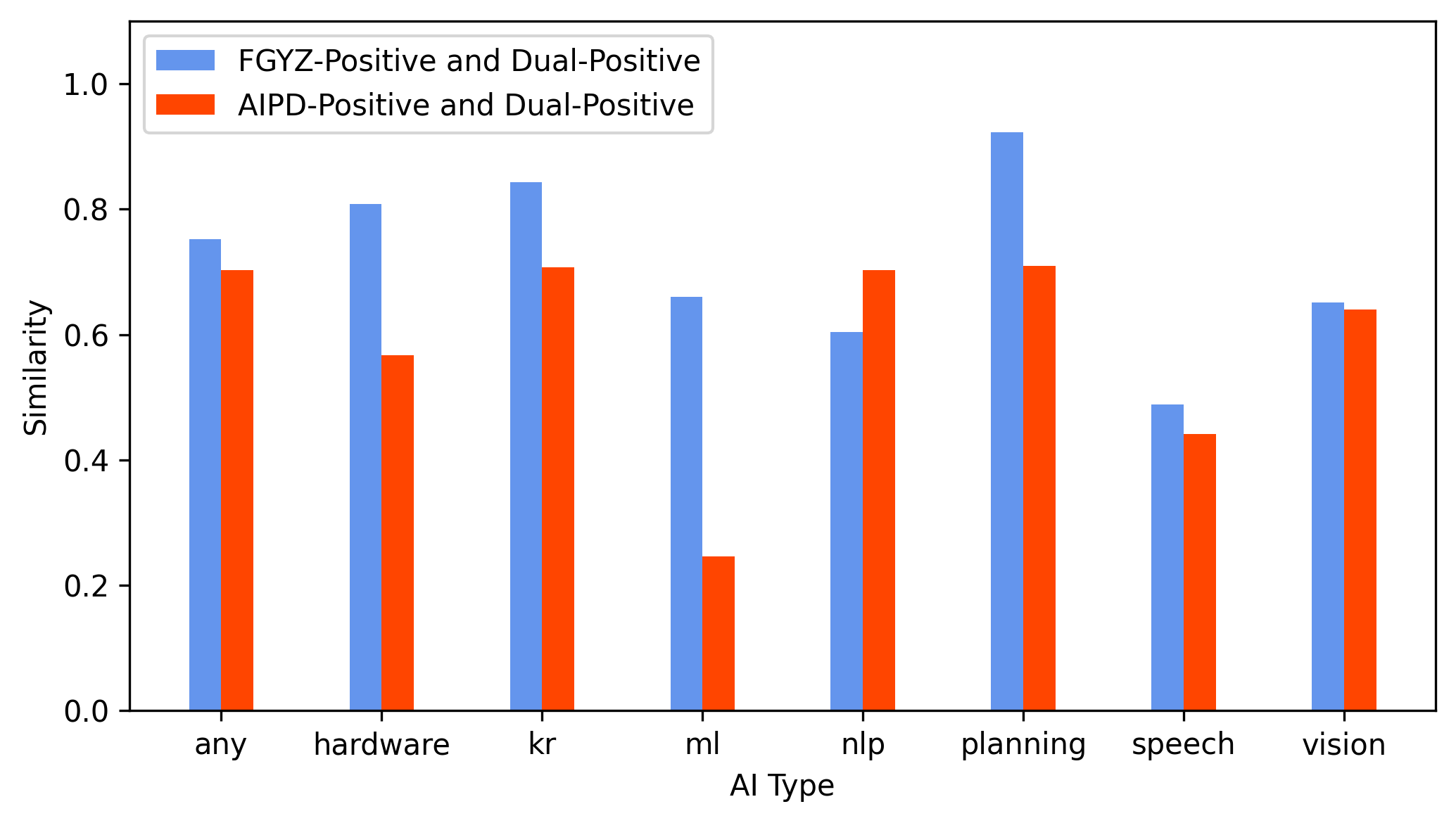}
\end{subfigure}

\vspace{2mm}

\begin{subfigure}[t]{\textwidth}
    \centering
    \caption{To Dual-Negative}
    \includegraphics[width=0.8\linewidth]{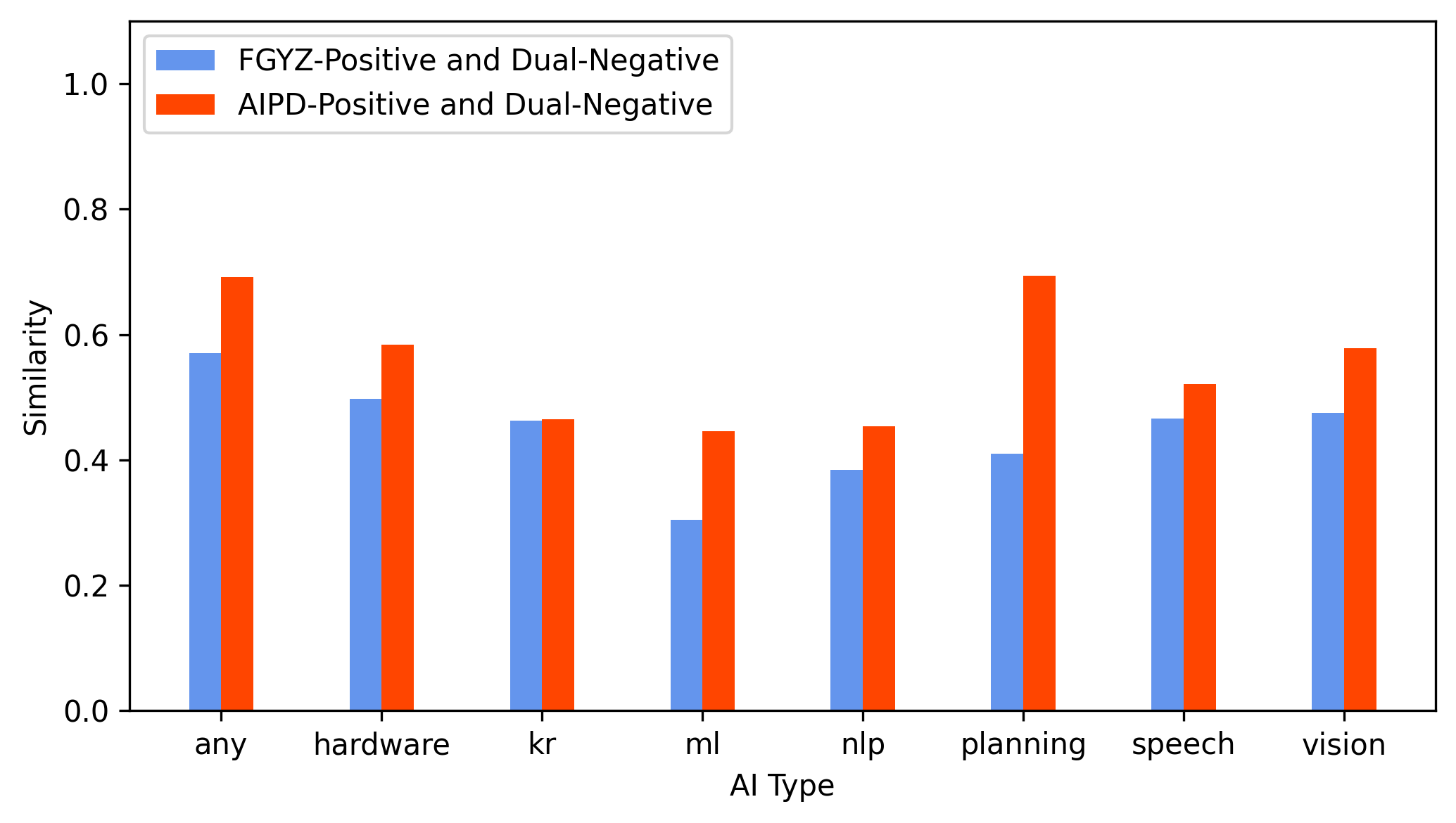}
\end{subfigure}
\end{figure}

Figure~\ref{fig:us_lexical_similarity} reports the lexical similarity between \textit{AIPD-Positive} and \textit{FGYZ-Positive} patents and the \textit{Dual-Positive} benchmark. Similarity is measured using the inner product of TF–IDF–weighted word distributions, so higher values indicate closer alignment in vocabulary usage with high-confidence AI patents. Across AI subfields—except for NLP—\textit{FGYZ-Positive} patents exhibit consistently higher lexical similarity to the Dual-Positive set than \textit{AIPD-Positive} patents. The differences are particularly pronounced in Machine Learning, Hardware, and Planning. The magnitude of these gaps indicates that patents uniquely identified by FGYZ employ technical language that more closely resembles that of high-confidence AI patents.

Overall, the lexical evidence complements the citation-based results. \textit{FGYZ-Positive} patents are not only more strongly connected to established AI patents through citations but also display greater similarity in technical vocabulary, providing additional support for the higher precision of the FGYZ classification.

\begin{center}
    [Insert Figure~\ref{fig:us_lexical_similarity} About Here]
\end{center}

\section{Extending Classification to Chinese Patents}\label{sec:extension-china}

Having established the internal validity of our classifier within the US patent system, we next examine the external validity: can a model trained on US data be applied reliably to Chinese patents? This question is central to our comparative analysis. To evaluate the performance of the FGYZ classifier--which was trained using US dataset--in the Chinese context, we develop a validation framework based on three considerations.

First, \textit{Linguistic Consistency}: Chinese patent records in our dataset include standardized English-language abstracts and claims, ensuring that the model operates in a linguistically consistent feature space despite the geographical difference. Second, \textit{Citational Connectivity}: We draw on the global nature of knowledge diffusion. If our model correctly identifies Chinese AI patents, these patents should display strong citation linkages to the established US AI frontier. Third, \textit{Semantic Homology}: we conduct a textual similarity check to verify that the underlying technical vocabulary remains comparable across the two systems, providing an intuitive validation of semantic consistency. 

To implement this, we partition the combined corpus into four mutually exclusive groups according to the classifications by our FGYZ classifier: \textit{Chinese-AI} and \textit{Chinese-NonAI}, against the \textit{US-AI} and \textit{US-NonAI}.

\begin{figure}[htbp]
\centering
\caption{Model Performance on Chinese AI Patent Classification}
\label{fig:cn_pred_prob_boxplot}
\vspace{5mm}
\caption*{This figure presents the distribution of predicted probabilities $\hat p_i$ generated by our classifier, separately for patents classified as AI-related (\textit{Yes}) and non-AI (\textit{Not}), across eight AI subfields. For each subfield, the boxplots summarize the conditional distributions $\hat p_i \mid \hat y_i = 1$ and $\hat p_i \mid \hat y_i = 0$, where $\hat y_i$ denotes the predicted label assigned by the model. Within each subplot, ``Yes'' indicates that the patent is classified as belonging to the corresponding AI subcategory, while ``Not'' indicates that it is classified as outside that subcategory by the FGYZ classifier. Higher predicted probabilities in the \textit{Yes} group reflect stronger model confidence in identifying AI-related patents. The degree of separation between the \textit{Yes} and \textit{Not} distributions provides a visual measure of classification sharpness and discrimination performance across subfields.}
\vspace{5mm}
\makebox[\textwidth][c]{
\includegraphics[width=1.1\textwidth]{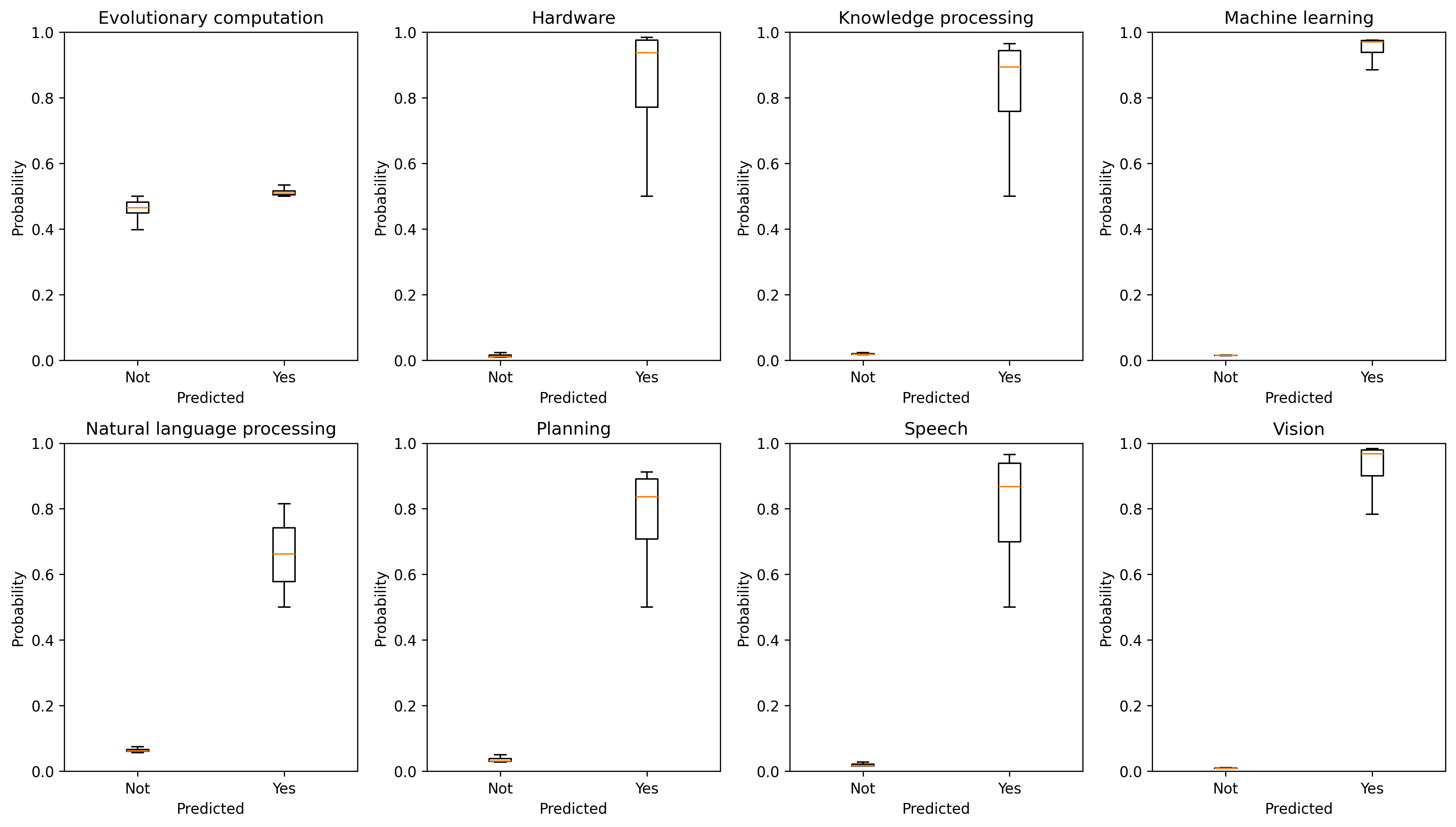}
}
\end{figure}

Before turning to the validation tests, we examine the distribution of predicted probabilities conditional on the \textit{predicted} label\footnote{Noted that there is a difference from Figure~\ref{fig:model_performance_visual} using the \textit{true} label as the x-axis to confirm our classifier showing sharp polarization on test data, we use \textit{predicted} label here to demonstrate that the model also has clear classification boundaries when applied to Chinese patent data.} across AI subfields, using the full set of Chinese patents (Figure~\ref{fig:cn_pred_prob_boxplot}). With the exception of the \textit{evolutionary computation} category, the predicted probabilities for AI and non-AI patents are clearly separated across subfields, suggesting that the model effectively distinguishes AI-related content. The degree of separation varies across subfields: more established and well-defined domains such as machine learning, computer vision, and speech recognition exhibit tighter distributions and higher classification confidence.
\begin{center}
    [Insert Figure~\ref{fig:cn_pred_prob_boxplot} About Here]
\end{center}

\subsection{Cross-Border Citation Validation}

We first examine whether Chinese AI patents are integrated into the global AI citation network. Using the citation-connectivity metric introduced in Equations (\ref{eq:R_A_to_B}) and (\ref{eq:C_AB}), we compute the preference for citation between groups, but restrict attention to only cross-country citations. Specifically, when evaluating Chinese patents, we consider only their citations to US patents and test whether \textit{Chinese AI} patents cite \textit{US AI} patents more intensively than \textit{US NonAI} patents. We apply the same approach to citations originating from the US. These cross-country connectivity patterns allow us to examine whether Chinese AI patents are more closely connected to the US AI technologies than to US non-AI technologies, thereby providing evidence on the model's validity beyond the country on whose data it was trained.

\begin{figure}[htbp]
\centering
\caption{Citation-Based Connectivity of Chinese and US Patents}
\label{fig:cn_mutual_cite_with_us}
\vspace{2mm}
\caption*{
This figure shows the citation-based connectivity between FGYZ-classified Chinese patents and US patents. Panel (a) reports the connectivity of Chinese AI and Chinese Non-AI patents to the US-AI group, while Panel (b) reports the corresponding connectivity to the US Non-AI group. Connectivity is computed using the bidirectional citation preference measure defined in Section~\ref{sec:connectivity}, restricted to cross-country citations from China to the US. Higher values indicate stronger technological relatedness. Across all AI subfields, Chinese-AI patents exhibit substantially stronger connectivity to US-AI patents and weaker connectivity to US Non-AI patents than Chinese Non-AI patents, providing evidence that the FGYZ classifier identifies Chinese AI patents that are more consistent with the technological core of AI innovation in the US.
}
\vspace{2mm}
\begin{subfigure}[t]{\textwidth}
    \centering
    \caption{To US AI Patents}
    \includegraphics[width=0.8\linewidth]{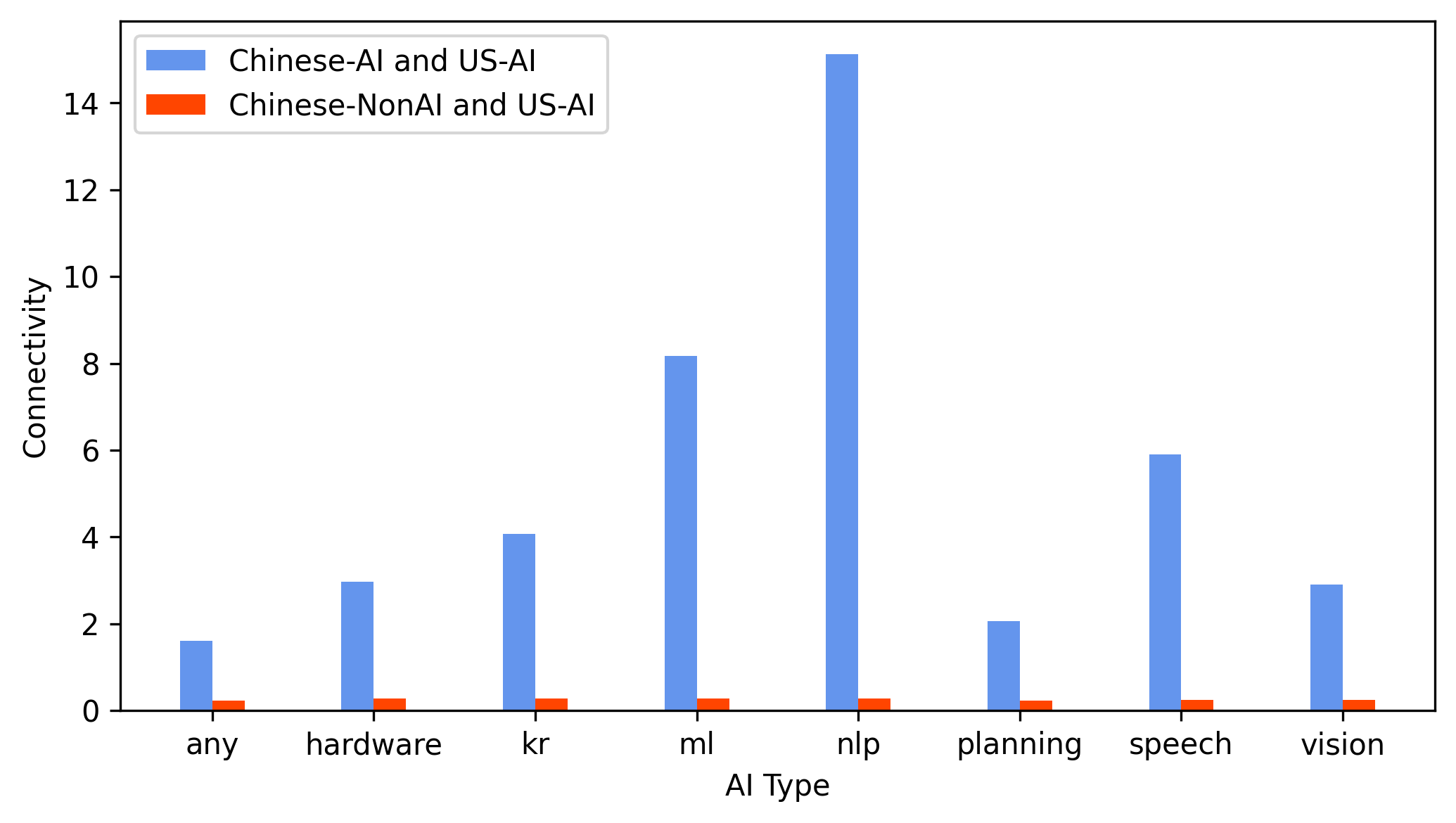}
\end{subfigure}

\vspace{2mm}

\begin{subfigure}[t]{\textwidth}
    \centering
    \caption{To US Non-AI Patents}
    \includegraphics[width=0.8\linewidth]{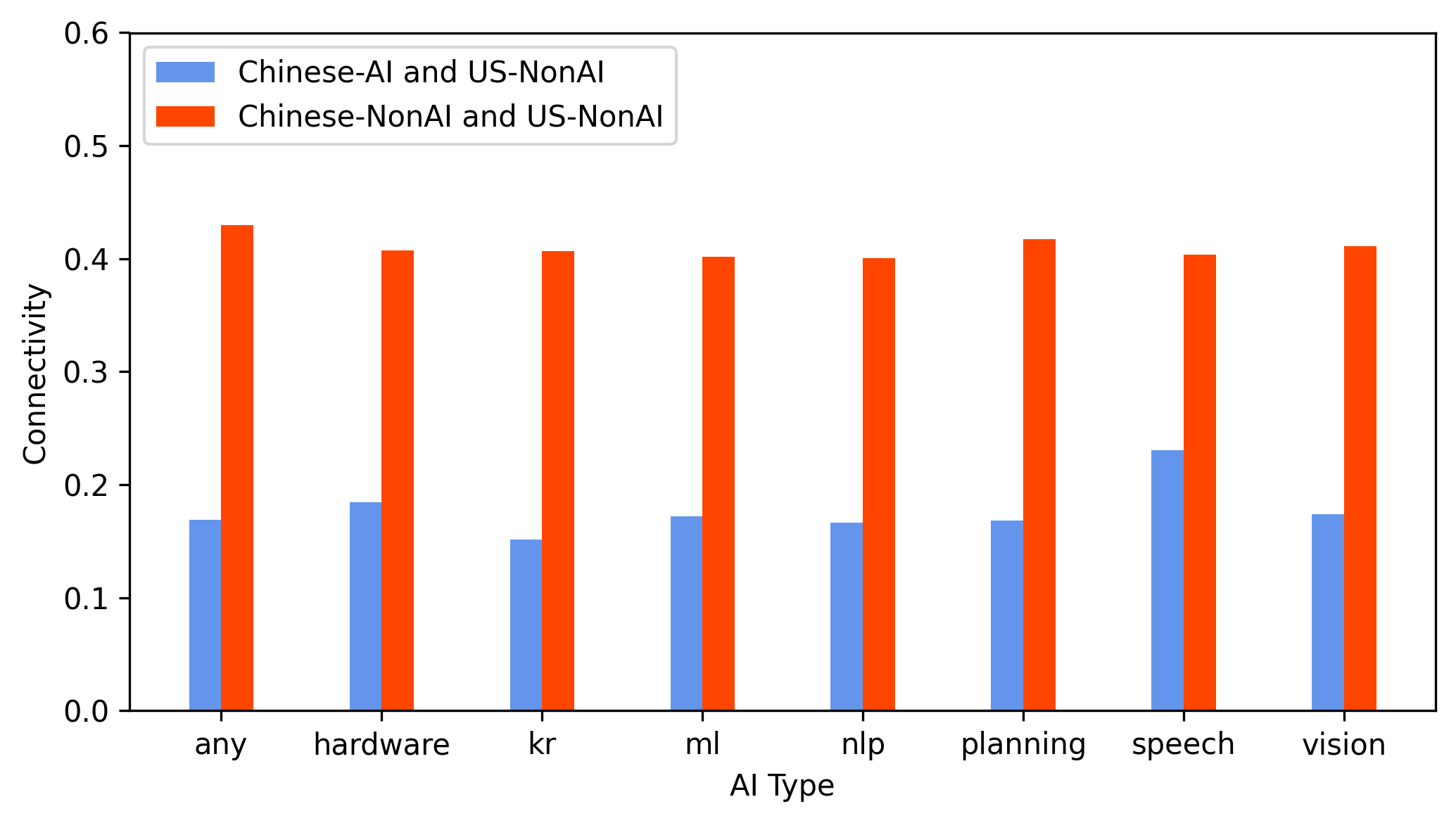}
\end{subfigure}
\end{figure}

Figure~\ref{fig:cn_mutual_cite_with_us} reports citation-based connectivity between FGYZ-classified Chinese and US patents, focusing exclusively on cross-country citations from China to the US. Panel (a) shows connectivity to US AI patents, while Panel (b) reports connectivity to US non-AI patents. Panel (a) reveals a clear difference between Chinese AI and Chinese non-AI patents. Across all AI subfields, Chinese AI patents show substantially stronger citation connectivity to US AI patents than Chinese non-AI patents. This difference is particularly pronounced in core AI domains such as natural language processing, machine learning, and speech, where the connectivity of Chinese AI patents exceeds that of Chinese non-AI patents by several multiples. These patterns suggest that Chinese patents classified as AI by FGYZ are tightly linked, through citation ties, to the US AI frontier.

Panel (b) presents the pattern for connectivity to US NonAI patents, where Chinese AI patents show consistently weaker connectivity than Chinese non-AI patents across all AI subfields, suggesting that Chinese AI patents have limited citation exposure to US non-AI technologies, whereas Chinese non-AI patents show stronger links to the US non-AI patents.

Taken together, the two panels provide strong cross-country validation evidence. Chinese patents identified as AI by FGYZ are more closely connected to US AI patents and less connected to US non-AI patents than their non-AI counterparts. 

\begin{center}
    [Insert Figure~\ref{fig:cn_mutual_cite_with_us} About Here]
\end{center}

\subsection{Cross-Border Lexical Validation}

In parallel, we calculate lexical similarity between Chinese and US patent groups using the TF--IDF--weighted distributional similarity measure described in Equation (\ref{eq:similarity_inner_product}). We construct the background distribution using the \textit{US-NonAI} group and compute the reweighted word distributions for \textit{Chinese-AI}, \textit{Chinese-NonAI}, and \textit{US-AI} patents. We then calculate the inner product between these distributions to quantify the extent to which Chinese AI patents resemble US AI patents in their vocabulary profiles. Together, citation-based and lexical analyses provide complementary evidence on whether FGYZ produces consistent and credible AI classifications for patents outside the US training domain.

\begin{figure}[htbp]
\centering
\caption{Lexical Similarity of Chinese and US Patents}
\label{fig:cn_lexical_similarity_with_us}
\vspace{2mm}
\caption*{
This figure shows the lexical similarity between FGYZ-classified Chinese and US patents. Similarity is computed as the inner product between the TF--IDF--weighted word distribution of each Chinese patent group and that of the FGYZ-identified US patents, following the procedure described in Section~\ref{sec:lexical_similarity}. Higher values indicate closer lexical alignment with the vocabulary profile of US patents. Across all AI subfields, Panel (a) shows that Chinese AI patents display substantially stronger lexical similarity to US AI patents than Chinese Non-AI patents, while Panel (b) shows that Chinese AI patents display substantially lower lexical similarity to US-NonAI patents than Chinese Non-AI patents, providing additional evidence that the FGYZ classifier successfully identifies Chinese patents that are more closely tied to the technological core of US AI innovation.
}
\vspace{2mm}
\begin{subfigure}[t]{\textwidth}
    \centering
    \caption{To US AI Patents}
    \includegraphics[width=0.8\linewidth]{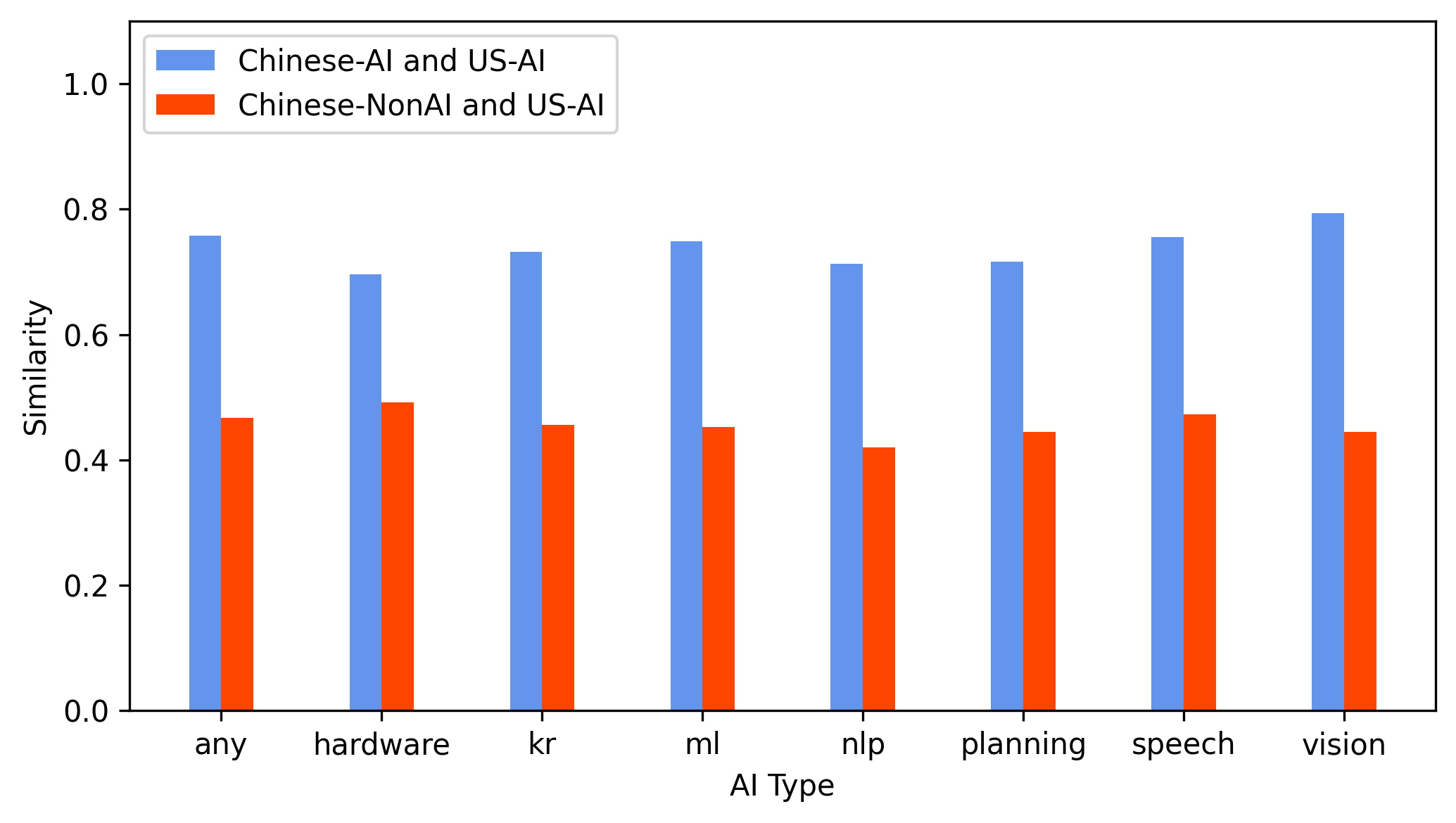}
\end{subfigure}

\vspace{2mm}

\begin{subfigure}[t]{\textwidth}
    \centering
    \caption{To US Non-AI Patents}
    \includegraphics[width=0.8\linewidth]{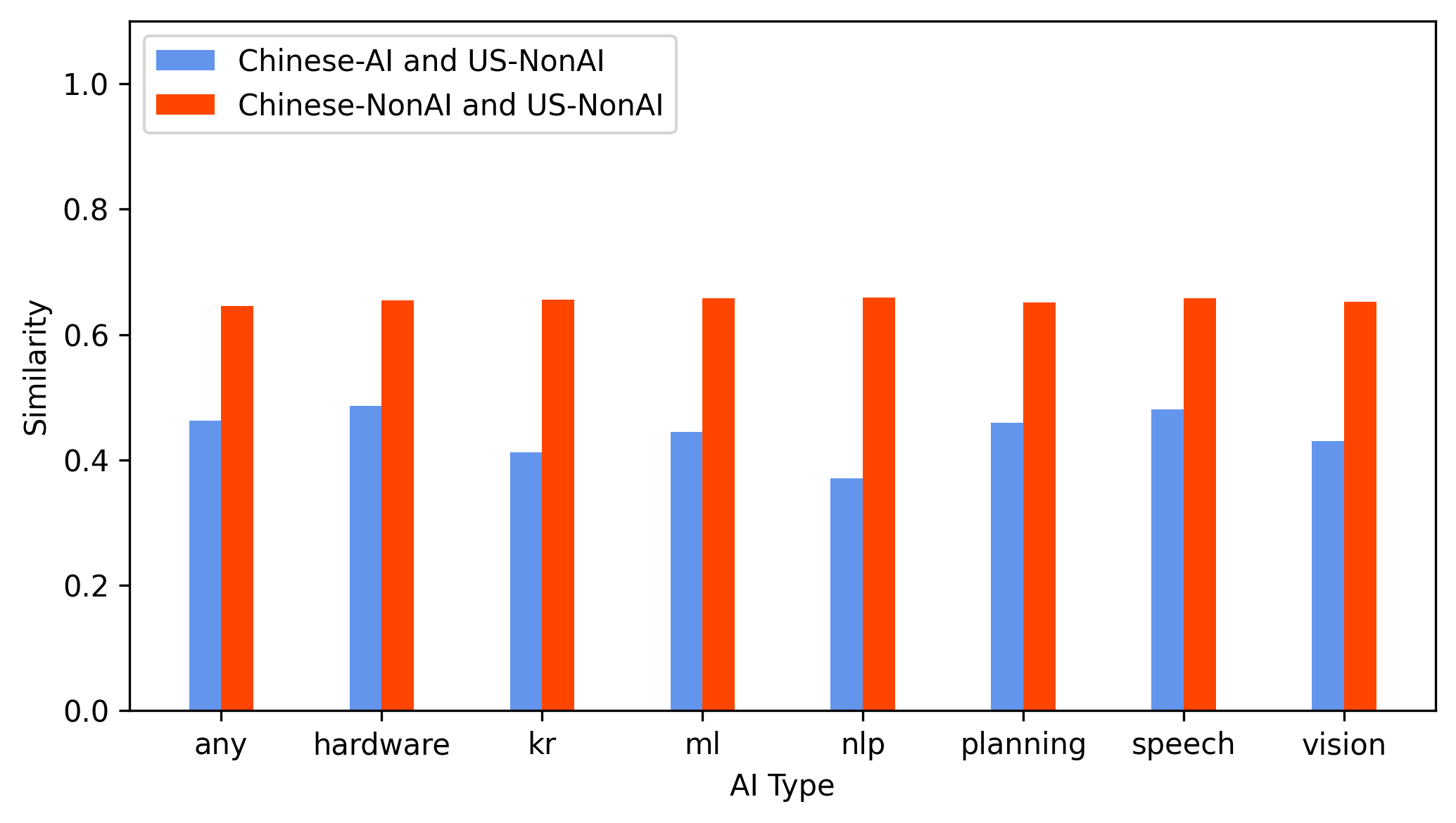}
\end{subfigure}

\end{figure}

Figure~\ref{fig:cn_lexical_similarity_with_us} reports the lexical similarity between FGYZ-classified Chinese patents and the FGYZ-identified US-AI benchmark. Across all AI subfields, Chinese AI patents show substantially higher lexical similarity to US AI patents than Chinese non-AI patents. The magnitude of the differences suggests that Chinese patents identified as AI by FGYZ use technical language that closely matches the vocabulary profile of US AI patents. 

\begin{center}
    [Insert Figure~\ref{fig:cn_lexical_similarity_with_us} About Here]
\end{center}

\section{Comparing US and Chinese AI Patents}\label{sec:facts}
The results above provide strong validation out-of-sample for the classifier. Beyond citation linkages, Chinese AI patents also resemble US AI patents in their lexical content, suggesting that the FGYZ classifier successfully identifies Chinese patents that align with the technological core of AI innovation in the US. We now document some stylized facts about the comparison of US and Chinese AI patents as classified by FGYZ.
\subsection{The Rise of AI Patents}
\begin{figure}[htbp]
\centering
\caption{Comparing AI Patents: USPTO vs. CNIPA}
\label{fig:AI_patents_both_country}
\vspace{2mm}
\caption*{This figure presents the number and percentage of AI patents each year, granted by the USPTO and the CNIPA. We define AI patents as those assigned by our model to at least one of the seven core AI subcategories: \textit{Machine Learning} (ML), \textit{Natural Language Processing}  (NLP), \textit{Speech}, \textit{Vision}, \textit{Planning}, \textit{Knowledge Processing} (KR), and \textit{Hardware}. Panel (a) shows the number of AI patents granted by the USPTO and the CNIPA each year, and Panel (b) reports the corresponding percentage of AI patents relative to all patents granted by each office.}
\vspace{2mm}
\begin{subfigure}[t]{\textwidth}
    \centering
    \caption{Number of AI Patents Granted by USPTO and CNIPA}
    \includegraphics[width=0.7\linewidth, height=8.5cm]{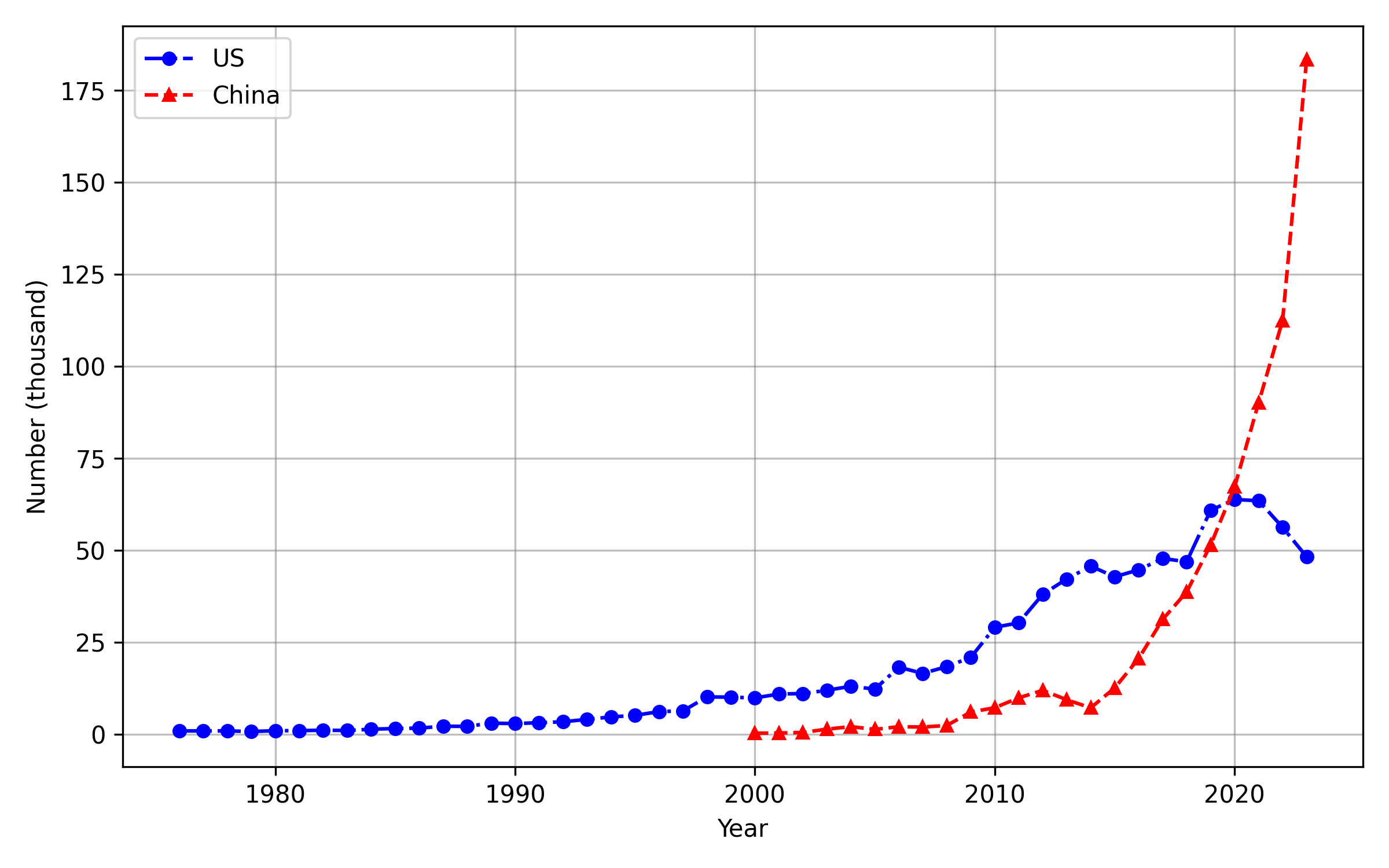}
\end{subfigure}

\vspace{2mm}

\begin{subfigure}[t]{\textwidth}
    \centering
    \caption{Percentage of AI Patents Among All Patents Granted by USPTO and CNIPA}
    \includegraphics[width=0.7\linewidth, height=8.5cm]{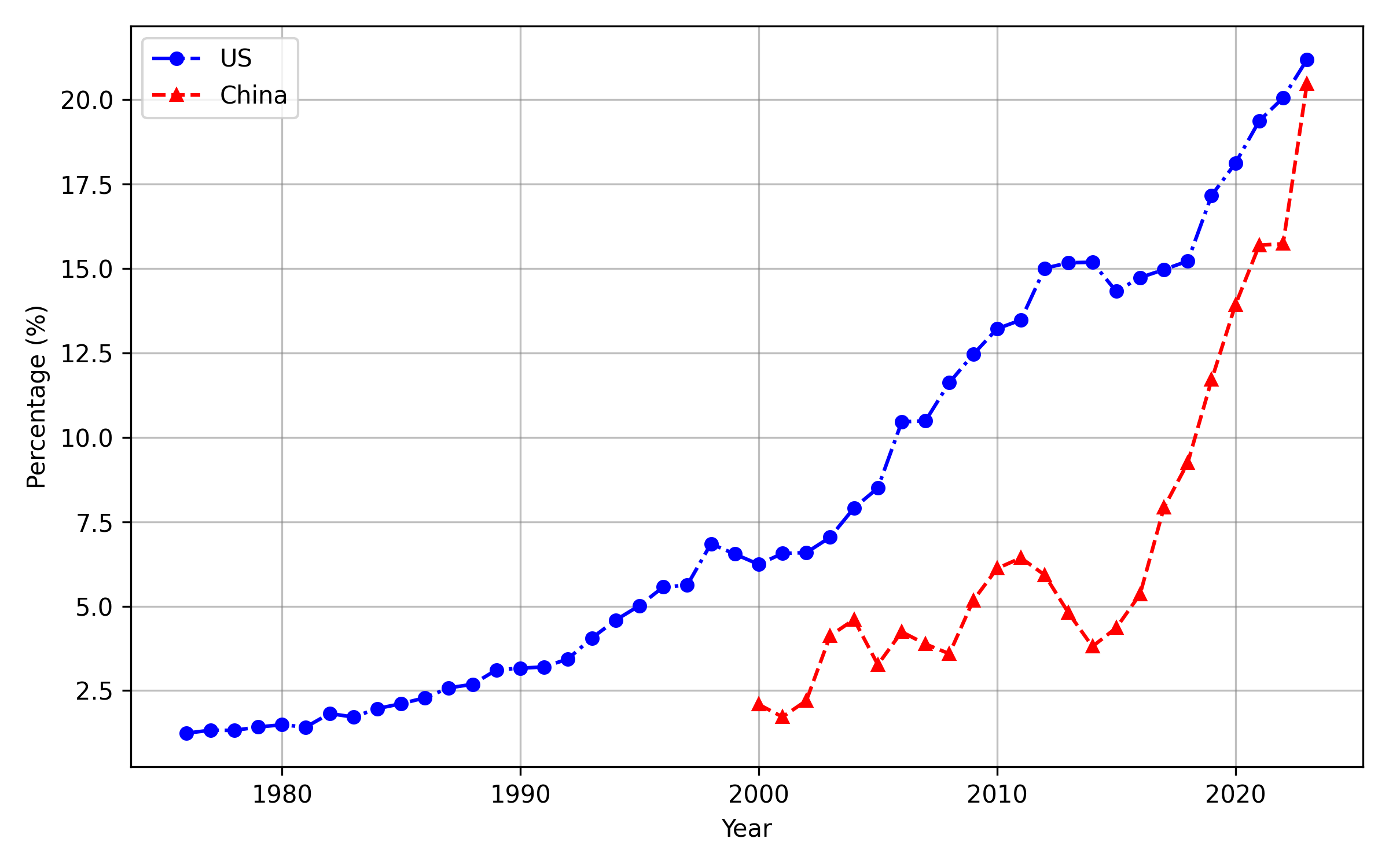}
\end{subfigure}
\end{figure}

Figure~\ref{fig:AI_patents_both_country} presents the aggregate trends in AI patenting activities in the US and China. Panel A shows the total number of AI patents, as classified by FGYZ, granted by the USPTO and CNIPA, and Panel B reports the share of AI patents as a percentage of the total patents granted by each office. From 2010 to 2023, the USPTO and CNIPA granted 659,504 and 651,629 AI patents, respectively. The plots show that both countries have experienced rapid growth in AI patenting, with a particularly sharp acceleration in recent years. In particular, China has overtaken the US in the total annual number of AI patents granted since 2020. Panel B shows that the percentages of AI patents among all patents in both countries reached 20\% in 2023, while back in 2014 the percentage was 3\% for China and 15\% for the US.

\begin{figure}[htbp]
\centering
\caption{Number of AI patents by Subcategory: USPTO versus CNIPA}
\label{fig:numbers_of_AI_patents}
\vspace{2mm}
\caption*{This figure presents the number of AI patents granted by the USPTO and the CNIPA each year. We focus on seven main AI subcategories: \textit{Machine Learning} (ML), \textit{Natural Language Processing} (NLP), \textit{Speech}, \textit{Vision}, \textit{Planning}, \textit{Knowledge Processing} (KR), and \textit{Hardware}. Panel (a) reports the number of AI patents in each subcategory granted by the USPTO each year, and Panel (b) reports those for the CNIPA. All counts are reported in thousands.
}

\vspace{2mm}
\begin{subfigure}[t]{\textwidth}
    \centering
    \caption{Number of AI Patents Granted by USPTO, by Subcategory}
    \includegraphics[width=0.8\linewidth, height=9cm]{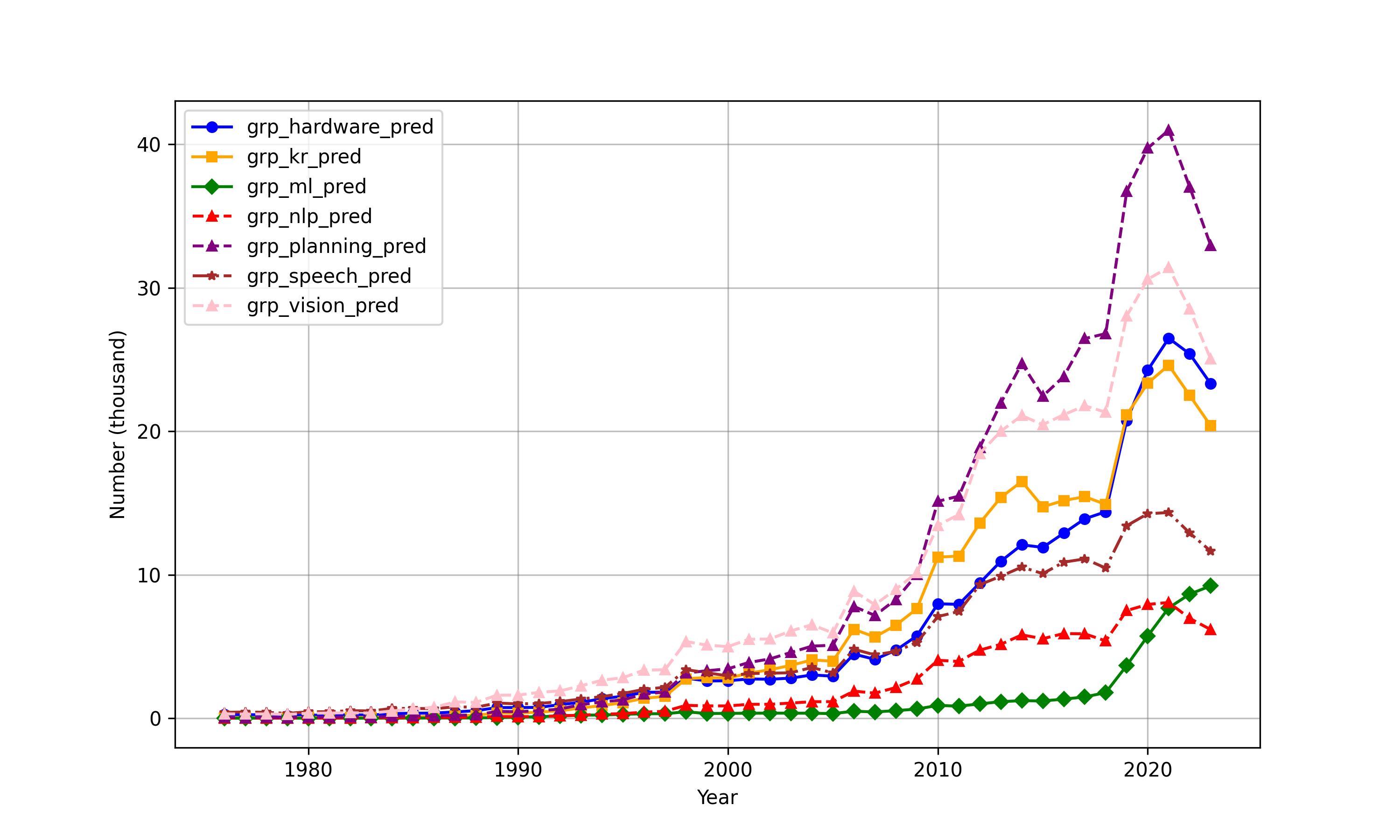}
\end{subfigure}

\vspace{2mm}

\begin{subfigure}[t]{\textwidth}
    \centering
    \caption{Number of AI Patents Granted by CNIPA, by Subcategory}
    \includegraphics[width=0.8\linewidth, height=9cm]{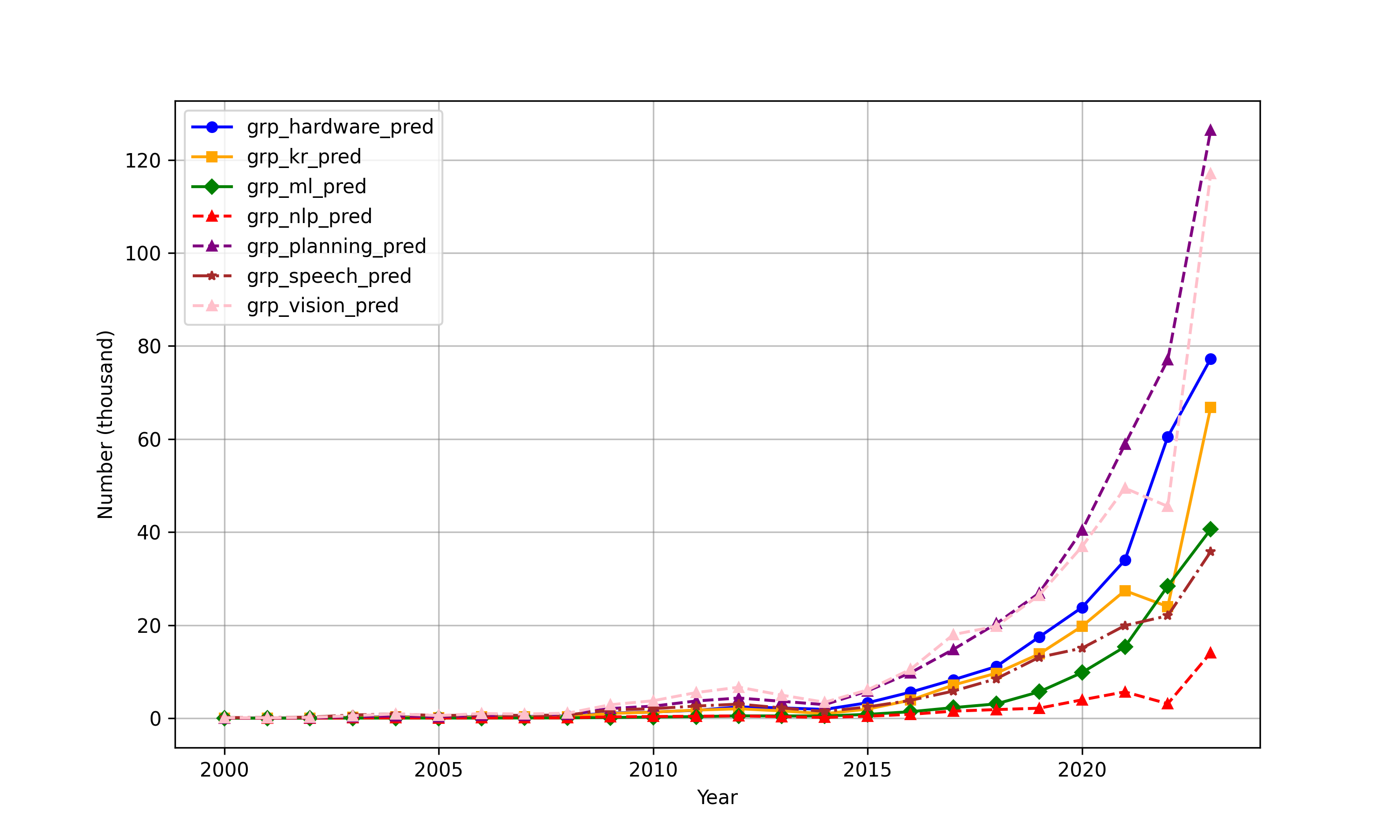}
\end{subfigure}
\hspace{0.04\textwidth}
\end{figure}

\begin{figure}[htbp]
\centering
\caption{Percentage of AI patents by Subcategory: USPTO vs. CNIPA}
\label{fig:proportion_of_AI_patents}
\vspace{2mm}
\caption*{This figure reports the percentage of AI patents among total patents granted by the USPTO and the CNIPA, respectively. We focus on seven main AI subcategories: \textit{Machine Learning}(ML), \textit{Natural Language Processing} (NLP), \textit{Speech}, \textit{Vision}, \textit{Planning}, \textit{Knowledge Processing} (KR), and \textit{Hardware}. Panel A presents the number of AI patents in each category granted by the USPTO each year, and Panel B presents those for the CNIPA.
}
\vspace{2mm}
\begin{subfigure}[t]{\textwidth}
    \centering
    \caption{Percentage of AI Patents by Subcategory Granted by USPTO, among All AI Patents}
    \includegraphics[width=0.8\linewidth, height=9cm]{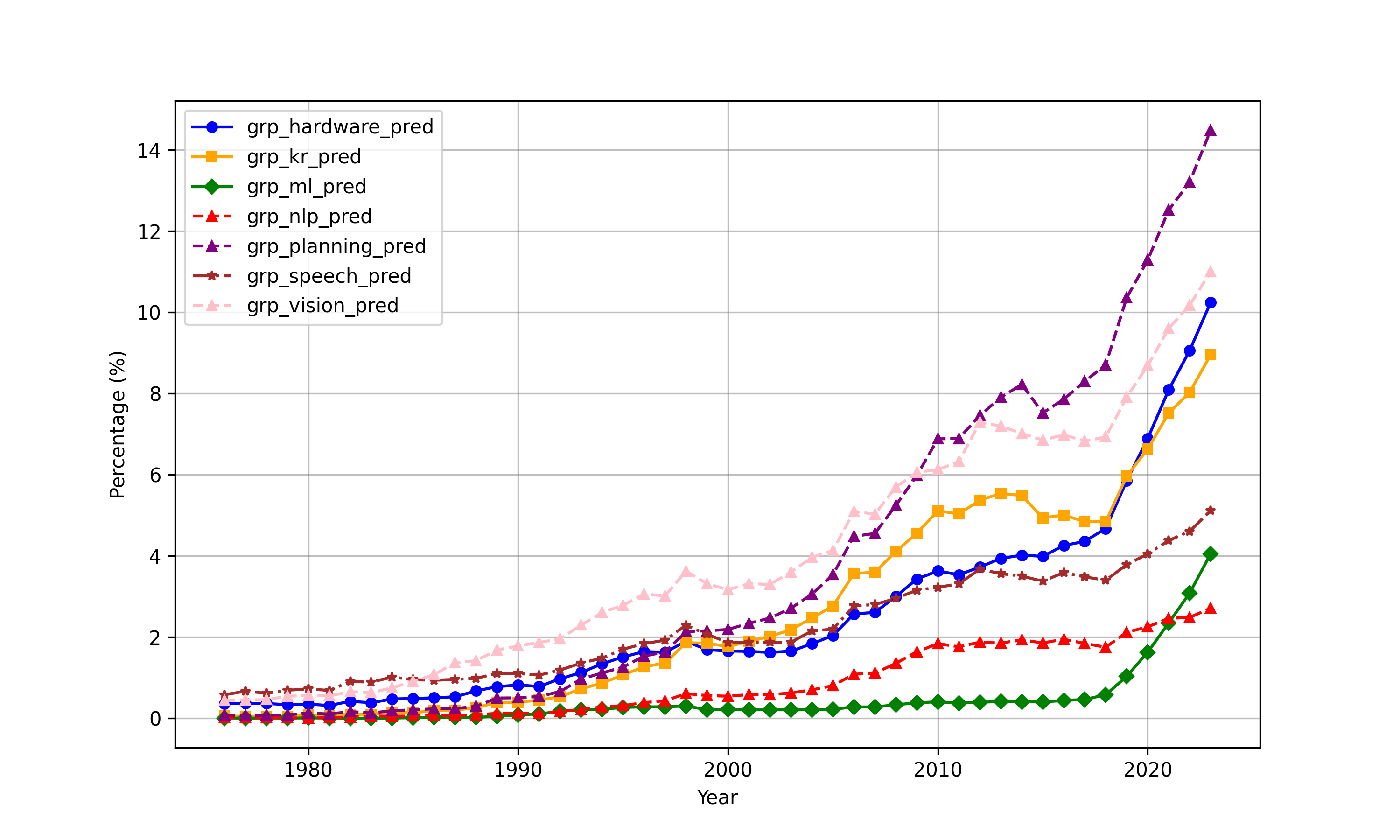}
\end{subfigure}

\vspace{2mm}

\begin{subfigure}[t]{\textwidth}
    \centering
    \caption{Percentage of AI Patents by Subcategory Granted by CNIPA, among All AI Patents}
    \includegraphics[width=0.8\linewidth, height=9cm]{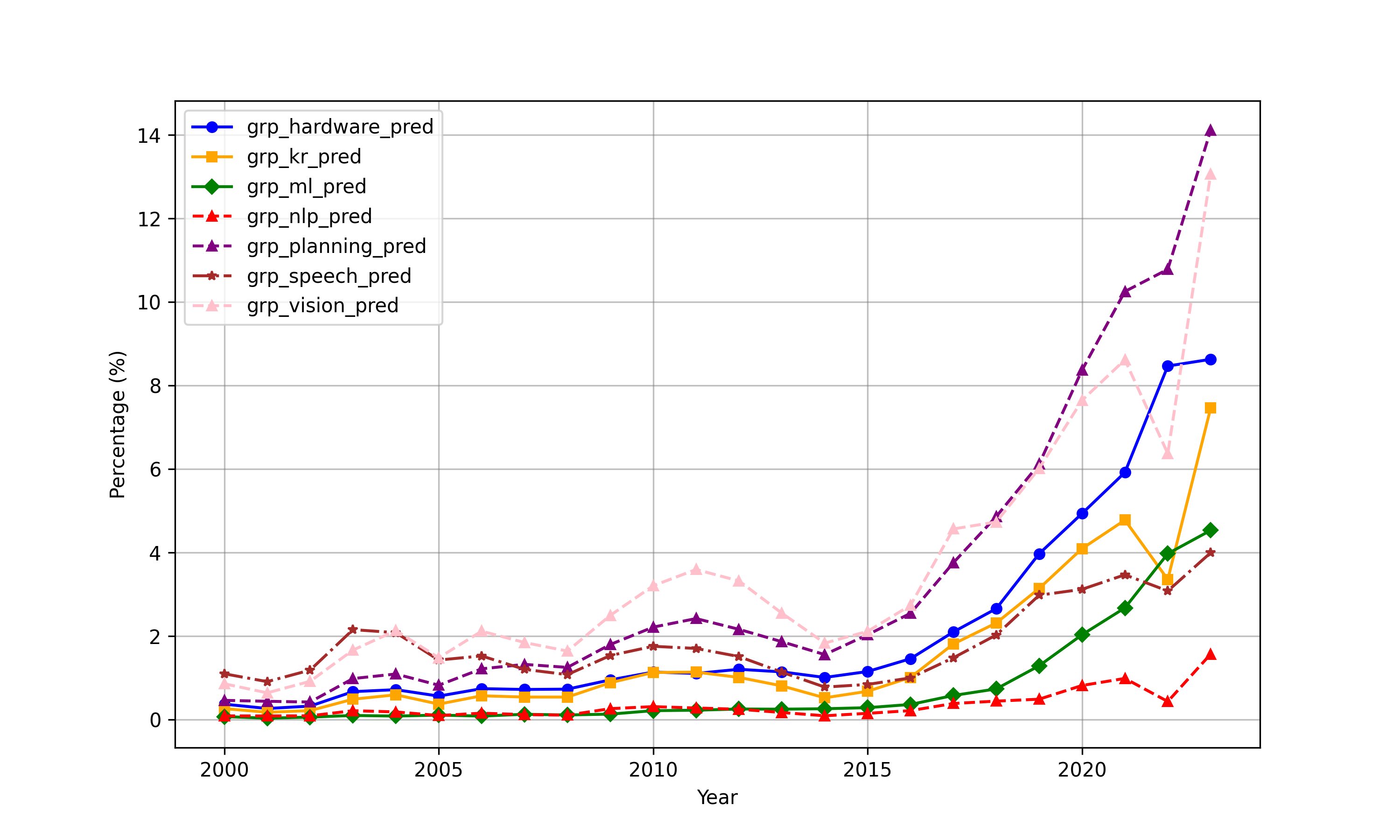}
\end{subfigure}

\end{figure}

Figures~\ref{fig:numbers_of_AI_patents} and~\ref{fig:proportion_of_AI_patents} present the annual number of AI patents and the percentage, by subcategory, of AI patents among all patents, granted by the USPTO and CNIPA, respectively.

\begin{center}
[Insert Figures~\ref{fig:AI_patents_both_country}-~\ref{fig:proportion_of_AI_patents} About Here]
\end{center}

Figures~\ref{fig:AI_patents_both_country} to~\ref{fig:proportion_of_AI_patents}  suggest several patterns in the evolution of AI innovation and patenting in the US and China. First, the total number of AI-related patents granted in China exceeds that of the US by approximately a factor of two, consistent with the broader trend in overall patenting activities in the two countries. For example, in 2023, China granted 183,302 AI patents compared to 48,197 AI patents granted in the US. In comparison, the total number of patents (AI and non-AI) is 895,488 in China and 227,499 in the US. The percentage of AI patents among all patents is 20.5\% in China and 21.2\% in the US. In general, the percentage of AI patents is at a comparable level and has converged in the two countries over the sample period 2010-2023. The mean percentage of AI patents is 16.2\% in the US and 9.4\% in China. However, one notable change since 2020 is the decline in overall US patenting activity, affecting both AI-related and non-AI patents. 

Second, the US and China have a similar distribution of AI patenting activity across the seven fields.  In both countries, the \textit{Planning} subcategory represents the largest share, representing approximately 14\% of the total patents. This is followed by patenting in \textit{Vision}, \textit{Hardware}, \textit{Knowledge Processing}, \textit{Machine Learning}, \textit{Speech}, and, to a lesser extent, \textit{Natural Language Processing (NLP)}.

Third, patterns of NLP-related patenting reveal a divergence between the US and China, particularly in recent years. In the US, NLP patenting has increased steadily over the past two decades, rising from negligible levels in the early 2000s to 7{,}930 patents in 2020, followed by a moderate decline thereafter. In contrast, China showed only relatively limited growth in NLP patenting prior to 2020, with a significant acceleration afterward. This trend coincides with the timing of US-led advances in large language models (LLMs). Taken together, the evidence suggests that while the US maintains a clear advantage in NLP technologies, the broader landscape of AI innovation reflects an intensifying competition between the two countries.

Fourth, we also observe a significant increase in China's patenting activity related to \textit{vision} technology beginning in 2022. This recent trend contrasts with the US, where related patenting activity has remained relatively stable over the same period. Both countries have historically treated computer vision as a core AI domain, consistently ranked among the top three subcategories by patent volume. However, the post-2022 acceleration in China may reflect increased industrial application and investment, or targeted policy support. This trend is consistent with the Chinese government’s broader strategic focus on applied AI, particularly in sectors such as surveillance, autonomous vehicles, and smart manufacturing, where vision technologies are critical \citep{beraja2023data}.

\subsection{Spatial Dynamics of AI Innovation}
\label{sec:spatial_dynamics}

To characterize the geography of the AI race, we first map the spatial evolution of inventive activity. For each patent in our sample, we geocode the assignee's location and aggregate activity into four six-year periods: 2000--2005, 2006--2011, 2012--2017, and 2018--2023. We construct kernel density estimates for both nations to visualize the transition from initial agglomeration to subsequent expansion. To ensure inter-temporal comparability, the density surfaces are normalized such that intensity levels reflect relative concentration rather than aggregate volume. This allows us to trace the \textit{extensive margin} of the AI economy—highlighting the emergence of new clusters and the saturation of established ones.

Figure~\ref{fig:cn_us_spatial_dynamic} illustrates these evolutionary paths. Panel (a) reveals the persistent dominance of US coastal ``super-clusters.'' Early AI activity is almost exclusively confined to the San Francisco Bay Area and the Northeast Corridor. Over the subsequent two decades, while we observe a ``trickle-down'' diffusion to secondary hubs (e.g., Austin, Seattle), the core pioneer locations intensify rather than dissipate. Panel (b) documents a distinct trajectory for China. Although early activity is similarly hyper-concentrated in the Beijing-Tianjin-Hebei, Yangtze River Delta, and Pearl River Delta mega-regions, the diffusion process appears notably more dynamic. By the most recent period, significant clusters of innovation emerge in the provincial capitals of inner regions, consistent with state-led initiatives to spread R\&D capacity \citep{allen2024centralization}.

Overall, the maps suggest a shared baseline of strong agglomeration but hint at divergent velocities of spatial spread in the two countries. We formally quantify this divergence in the next section.

\begin{center}
    [Insert Figure~\ref{fig:cn_us_spatial_dynamic} About Here]
\end{center}

\subsubsection{Spatial Diffusion of AI Technology}

To measure how AI innovation spreads across space, we adopt the diffusion approach of \cite{Kalyani2025diffusion}. Rather than focusing on static measures of geographic concentration, this framework captures how quickly AI activity expands from its early centers to other locations.

We define \textit{pioneer hubs} as the places where AI activity first took hold—specifically, the ten locations with the largest cumulative stock of AI patents during the first five years of the sample. In the US, we measure location at the \textit{Core Based Statistical Area (CBSA)} level, which reflects integrated labor markets. In China, we use \textit{prefecture-level cities}, the main administrative unit for local industrial policy and economic development.\footnote{In the US, a CBSA groups nearby cities and counties that are closely linked through commuting. Hence, rather than treating each city or county as a separate location, a CBSA treats the entire commuting area as one unit, which better reflects where economic activity and innovation actually take place (for example, \textit{Silicon Valley}). In China, prefecture-level cities are administrative regions below the provincial level that include a central city and surrounding counties, and they are the main unit through which local economic and industrial policies are implemented.}

We then construct the \textit{Diffusion Share} ($D_t$) for each country and year $t$:
\begin{equation}
D_t = \frac{\sum_{i \notin \text{Pioneers}} \text{New Patents}_{i,t}}{\sum_{all} \text{New Patents}_{t}}
\end{equation}

This index measures the proportion of new AI inventions that come from locations outside of the original pioneer hubs. Intuitively, a higher value of $D_t$ means that more AI innovation is taking place beyond the early hubs, while a stable $D_t$ suggests that activity remains geographically concentrated.

\begin{figure}[htbp]
\centering
\caption{Spatial Diffusion of AI and Non-AI Patents}
\label{fig:cn_us_spatial_diffusion}
\vspace{2mm}
\caption*{
This figure plots the \textit{diffusion share} of AI and Non-AI patents over time, using the methodology of \cite{Kalyani2025diffusion}, defined as the fraction of newly granted AI patents in year $t$ that originate from locations outside the initial pioneer hubs. Pioneer hubs are identified as the top ten \textit{locations} with the highest cumulative number of AI patents during the first five years of the sample period, that are \textit{Core Based Statistical Areas(CBSA)-level} locations in 1976--1980 for the US and prefecture-level cities in 2000--2004 for China. For each country and each year, \textit{diffusion share} is calculated as the share of new (Non-)AI patents granted outside these pioneer hubs relative to total (Non-)AI patents granted in the respective country. A rising \textit{diffusion share} suggests increasing geographic spread of (Non-)AI innovation beyond early centers, whereas a flat diffusion share over time suggests persistent spatial concentration.}

\vspace{5mm}
\includegraphics[width=1.0\textwidth]{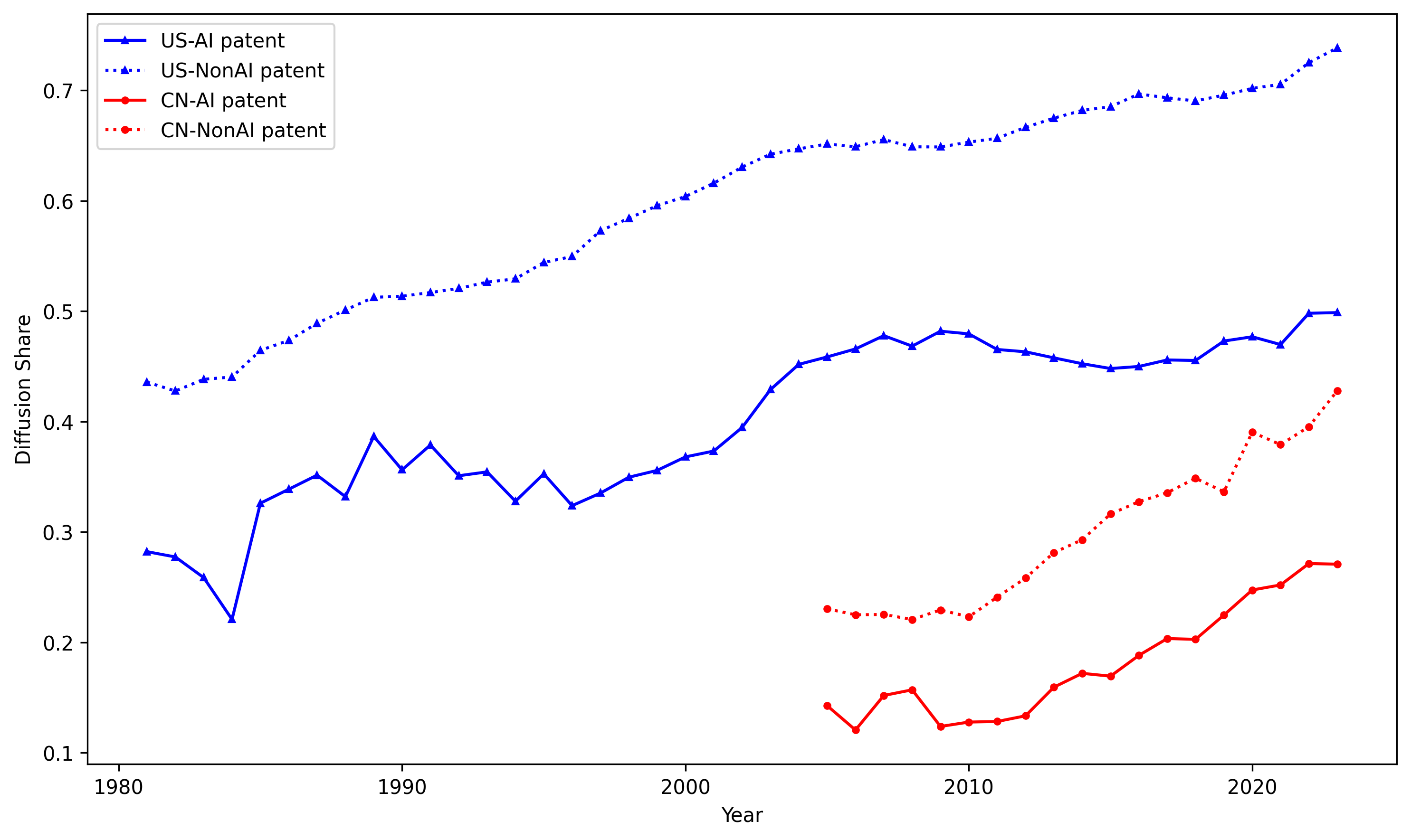}
\end{figure}

Figure~\ref{fig:cn_us_spatial_diffusion} plots the diffusion curves for the US and China, showing how AI activity spreads over time in the two countries. We observe three stylized facts. First, the two countries start from very different points. In China, AI innovation was initially highly concentrated: when our Chinese sample begins in 2005, locations outside the pioneer hubs account for only about 15\% of AI patents. The US, by contrast, was already more geographically dispersed. As early as 1980, non-pioneer locations produced nearly 28\% of AI patents, and this share had risen to about 35\% by 2005.

Second, the pace and stage of diffusion differ significantly between the two countries. Although China starts from a much lower base, AI activity spreads rapidly across locations: the share of AI patents from non-pioneer areas nearly doubles, from about 13\% to around 26\% over the sample period. This trend continues until the end of the sample. In contrast, while the US shows a higher level of geographic dispersion overall, the diffusion process slows down after 2010. The non-pioneer share increases from 28\% in 1981 to approximately 48\% in 2010 and remains nearly 50\% by 2023. The relatively flat curve between 2010 to 2023 suggests that further geographic expansion has largely leveled off. Together, these patterns suggest that AI activity in the US has reached a broadly distributed pattern, whereas China is still experiencing a rapid geographic expansion. 

Third, the figure also reports diffusion shares for non-AI patents. In both countries, AI patents consistently display lower diffusion shares than non-AI patents, suggesting that AI innovation is more geographically concentrated. The contrast is especially visible in the US: while non-AI innovation continues to spread gradually over time, AI diffusion rises through the mid-2000s and then levels off. In China, the diffusion of both AI and non-AI patents increased markedly after 2010, although AI remains less dispersed than non-AI. In general, AI activity appears to be more spatially concentrated than other forms of innovation, but the stage of diffusion differs significantly between the two countries.

\begin{center}
    [Insert Figure~\ref{fig:cn_us_spatial_diffusion} About Here]
\end{center}

\subsection{Top AI Patent Assignees}

To characterize the institutional structure of AI inventions in the US and China, we identify the leading assignees within each AI technology subfield. For every subfield, we rank assignees separately by country according to the number of AI patents identified using the FGYZ classification. We then retain the top ten firms or organizations in each subfield, showing the primary contributors to AI patenting activity in both countries. 

Table~\ref{tab:cn_us_assignee} reports the results. Panel A shows that AI patenting in the US is led by large multinational technology firms across all subfields. IBM and Microsoft consistently appear among the top assignees in every AI category, frequently occupying the first two positions. Other major firms—such as Google, Amazon, Samsung, and Intel—also rank highly in multiple subfields, reflecting broad technological portfolios rather than narrow specialization. Hardware- and vision-related AI patents show a stronger presence of electronics firms, including Canon, Sony, Toshiba, Fujitsu, while software- and data-intensive subfields such as machine learning, NLP, and planning are led primarily by platform and enterprise software firms. Overall, the US rankings suggest a relatively stable set of incumbent firms driving AI innovation across diverse technological areas.

Panel B suggests a more diverse institutional landscape in China. Although large technology firms such as Tencent and Baidu dominate most AI subfields—often ranking first or second, state-owned enterprises and universities play a much more visible role, particularly in hardware, planning, and vision. Firms such as Huawei and State Grid appear frequently across subfields, reflecting their broad engagement in applied AI technologies. In contrast to the US, Chinese universities, including Tsinghua University, Zhejiang University, UESTC, BUAA, rank among the top assignees in multiple subfields, suggesting the significant role of academic institutions in China’s AI patenting activity.

Taken together, this table illustrates significant differences between the two countries in the institutional landscape of AI inventions: AI patenting in the US is concentrated among a small group of diversified multinational firms, whereas China exhibits a more mixed structure involving private technology firms, state-owned enterprises, and universities, with variation across AI subfields.

\begin{center}
    [Insert Table~\ref{tab:cn_us_assignee} About Here]
\end{center}

\begin{figure}[htbp]
\centering
\caption{Innovation Concentration of AI and Non-AI Patents}
\label{fig:cn_us_assignee_concentration}
\caption*{
This figure plots the measures of assignee-level concentration for AI and non-AI patents over time, using the method in \cite{kwon2024100}. For each year, patents are divided into an AI sample and a non-AI sample consisting of all remaining patents. Within each sample, assignees are ranked by their annual patent counts. The concentration measures are the Top 1\% share, defined as the fraction of patents held by the top 1\% of assignees, and the Top 10 share, defined as the fraction of patents held by the ten largest assignees. Panel (a) reports the time series for the US and Panel (b) for China. Higher values indicate stronger winner-take-all dynamics and greater dominance by superstar firms.}

\begin{subfigure}[t]{\textwidth}
    \centering
    \caption{US Patents}
    \includegraphics[width=0.74\linewidth]{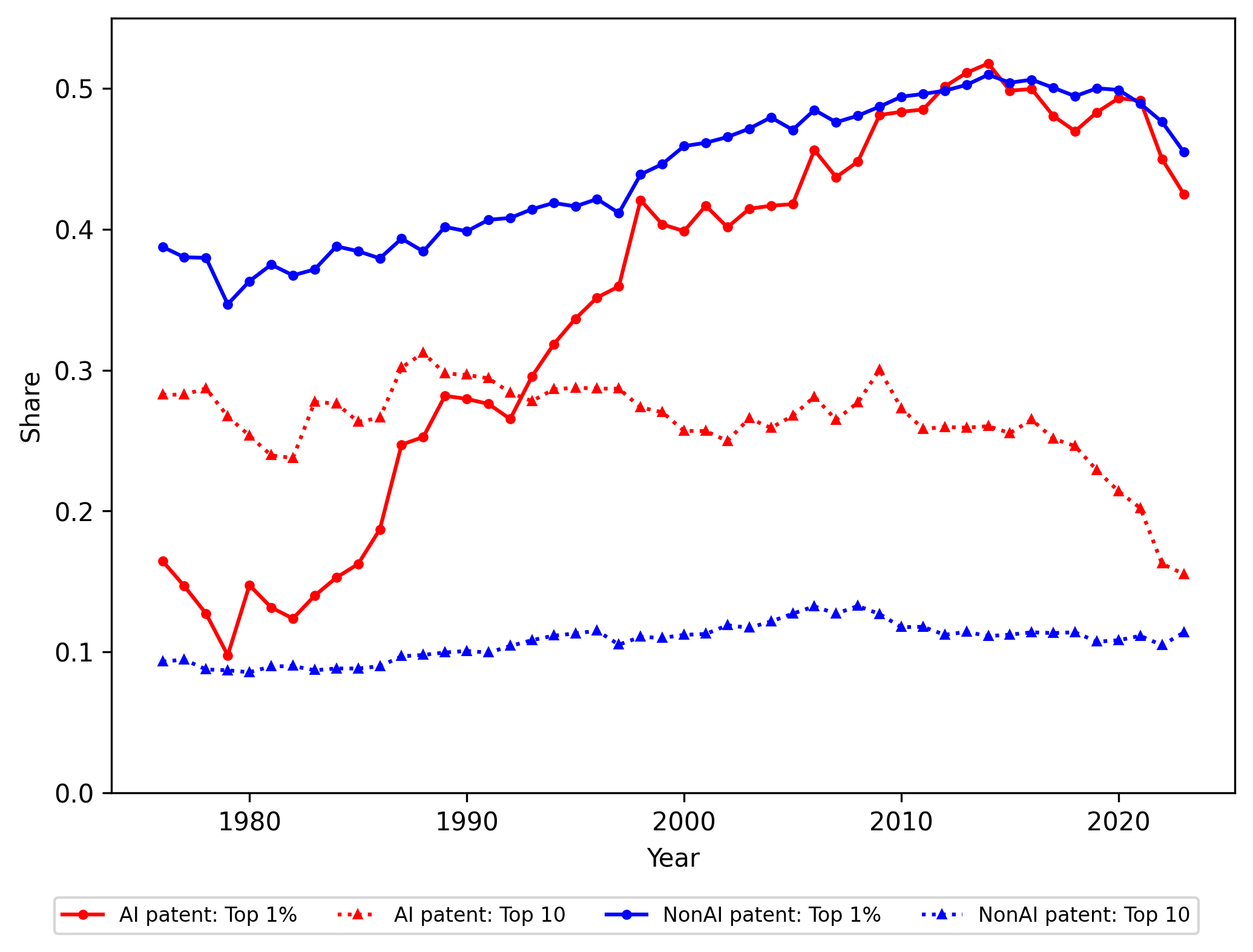}
\end{subfigure}

\vspace{2mm}

\begin{subfigure}[t]{\textwidth}
    \centering
    \caption{Chinese Patents}
    \includegraphics[width=0.74\linewidth]{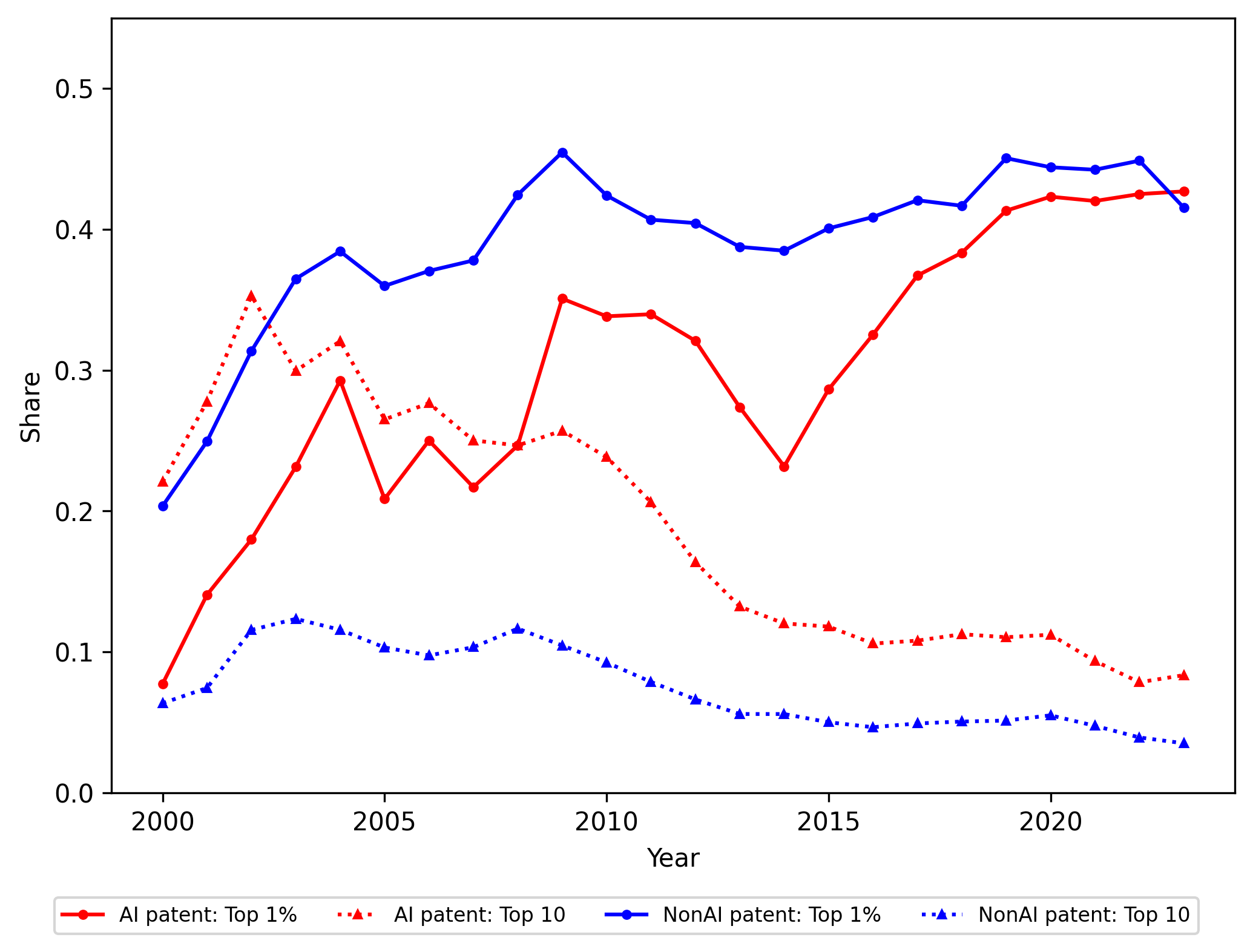}
\end{subfigure}
\end{figure}

Figure \ref{fig:cn_us_assignee_concentration} shows the dynamics of assignee concentration for the US and China, respectively. Panel A shows that AI patenting in the US has become increasingly concentrated over time, but not simply because the same few firms dominate throughout. From the early 1990s onward, the share of AI patents held by the top 1\% of assignees rises steadily and eventually exceeds the share held by the top ten firms. The share peaks at just over 50\% in the mid-2010s. This reflects the rapid expansion of the AI patenting landscape: as more firms become active, the top 1\% includes a wider group of leading companies rather than just the very largest incumbents. At the same time, the share accounted for by the top ten assignees declines slightly after around 2016, suggesting greater competition and turnover among the largest players. By contrast, concentration in non-AI patenting remains comparatively stable over time, with no similar divergence between the top-1\% and top-ten measures. These patterns point to stronger winner-take-most dynamics that are specific to US AI innovation. 

Panel B shows that AI patenting in China follows a pattern similar to that observed in the US. The share of AI patents held by the top 1\% of assignees rises over time, while the share held by the top ten assignees declines, suggesting increasing concentration alongside growing competition within the upper tier. As in the US, this divergence reflects an expanding pool of leading AI assignees rather than dominance by a fixed set of incumbents, and thus, also points to strong winner-take-most dynamics in Chinese AI innovation. What differs, is that, this expanding elite in China, as we have shown in Table~\ref{tab:cn_us_assignee}, reflects a more heterogeneous mix of private technology firms, state-owned enterprises, and universities. Similarly to the US, concentration in non-AI patenting also remains relatively flat over time, with no similar divergence between the two measures.

\begin{center}
    [Insert Figure~\ref{fig:cn_us_assignee_concentration} About Here]
\end{center}

\subsection{Economic Values of AI versus Non-AI Patents}

A central debate in the economics of innovation concerns the quality of China’s rapid patent expansion. Critics argue that government subsidies and administrative targets have inflated patent volumes with low-quality innovations that carry less commercial value \citep{fang2018corruption}. Hence, we examine the stock market valuation of Chinese AI patents and compare the market responses to those of US AI patents. If these patents serve mainly as instruments for obtaining government subsidies, their publication should attract less attention from investors. If, instead, they capture economically meaningful technological advances, their disclosure should be associated with positive abnormal stock-market returns and comparable to those observed for US AI patents, indicating that investors view them as value-relevant assets. 

We follow the approach of \cite{kogan2017technological}, which estimates the private economic value of a patent based on the firm's abnormal stock return within a narrow window around the grant date. This market-based identification strategy allows us to directly measure the expected future cash flows of the invention, and hence avoid the noise of citation-based measures.

For US patents, we directly use the patent-level value estimates reported in \cite{kogan2017technological}, which are constructed from a large panel of publicly listed US firms. For Chinese patents, we construct comparable value measures by applying the same methodology to publicly listed Chinese firms. This procedure yields patent-level value estimates that are conceptually aligned with the US benchmark and directly comparable across countries. Then we examine differences in the economic value of AI and non-AI patents within each country, as well as cross-country comparison in the relative value premium associated with AI innovation. 

\begin{figure}[htbp]
\centering
\caption{Economic Value of AI and Non-AI Patents}
\label{fig:cn_us_patent_value}
\vspace{2mm}
\caption*{
This figure shows the economic value of AI and non-AI patents in the US and China. Panel (a) reports the average market-based patent value for US\ listed firms in constant 1982 USD (millions), using CPI deflation and the estimates from \cite{kogan2017technological}. Panel (b) reports the corresponding values for Chinese listed firms in constant 1990 CNY (millions), also CPI deflated and constructed by applying the same approach based on abnormal stock returns around patent publication. In both countries, AI patents exhibit systematically higher economic value than non-AI patents across all seven AI technology subfields.
}
\begin{subfigure}[t]{\textwidth}
    \centering
    \caption{US Patents}
    \includegraphics[width=0.9\linewidth]{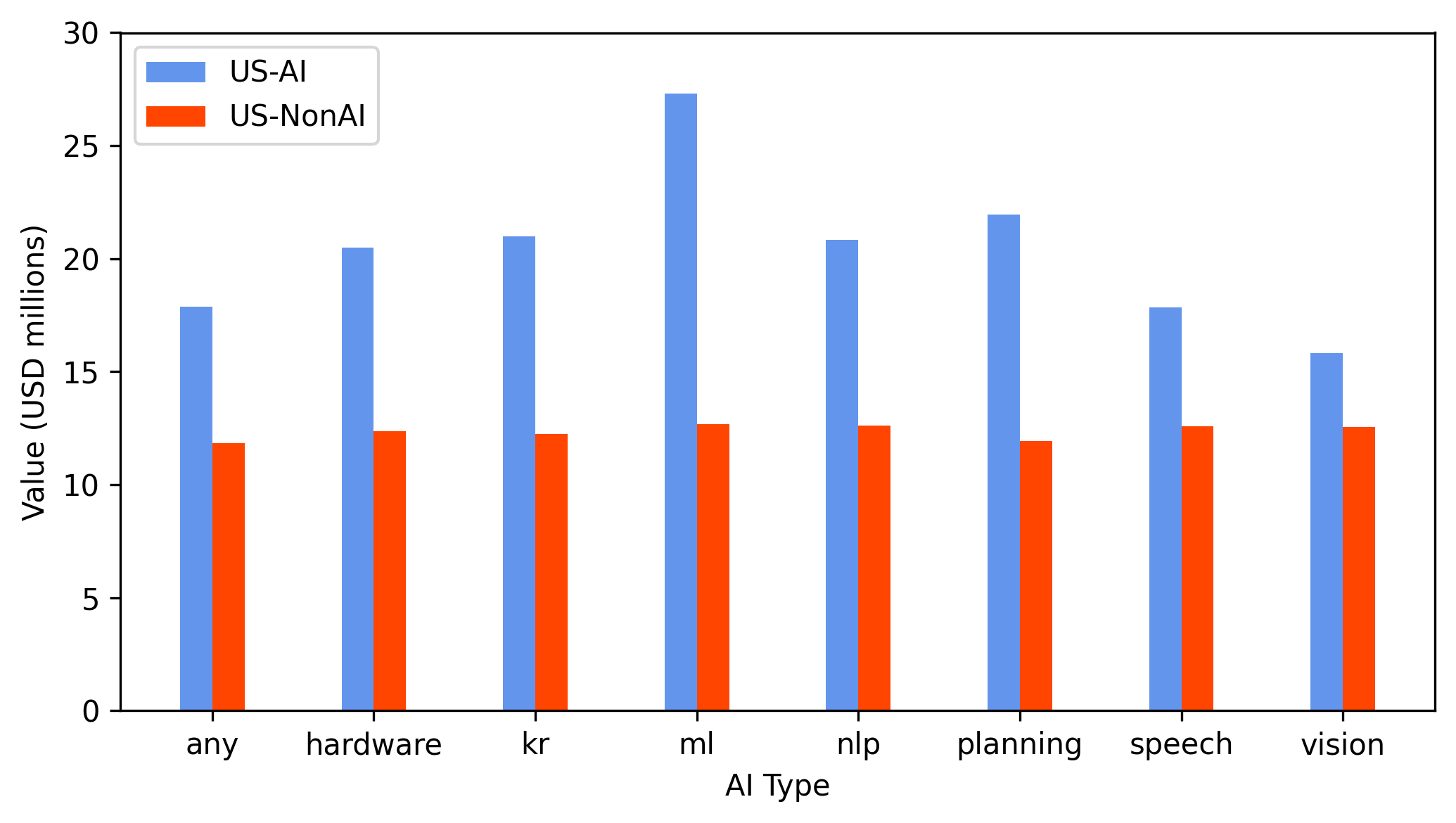}
\end{subfigure}

\vspace{2mm}

\begin{subfigure}[t]{\textwidth}
    \centering
    \caption{Chinese Patents}
    \includegraphics[width=0.9\linewidth]{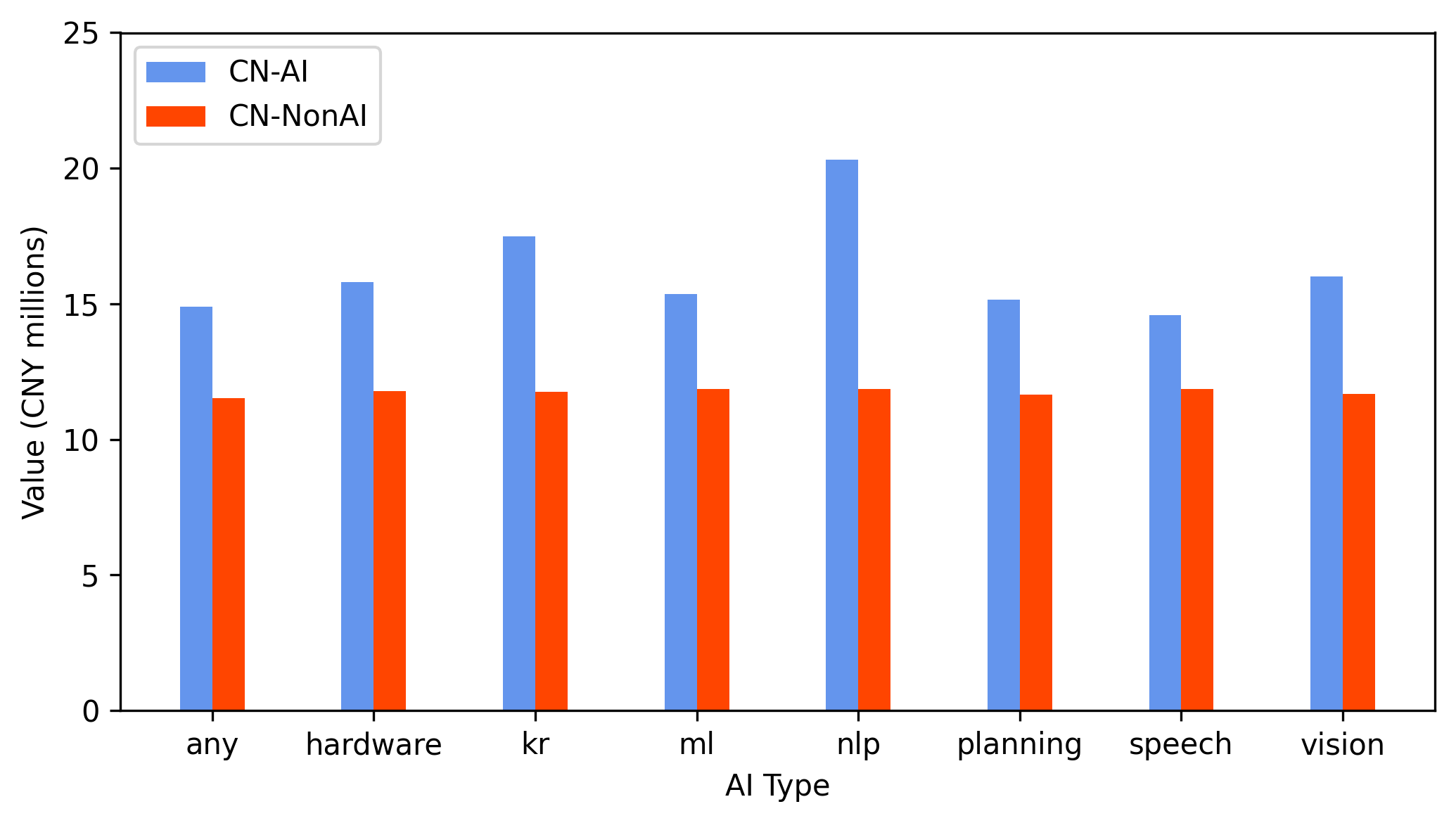}
\end{subfigure}
\end{figure}

Figure~\ref{fig:cn_us_patent_value} compares the average market-based economic value of AI and non-AI patents across AI technology subfields in the US (Panel A) and China (Panel B). In both countries, AI patents are consistently more valuable than non-AI patents across all subfields, suggesting a robust AI-related value premium. Panel A shows that, for US patents, the value difference between AI and non-AI patents is substantial across all subfields. The premium of AI patents is particularly pronounced in machine learning, planning, and NLP. Even in more hardware-oriented subfields such as vision and speech, AI patents are associated with higher market valuations, although the gap is smaller than in software-intensive domains.

Panel B presents comparable patterns for Chinese patents. AI patents held by Chinese listed firms also show a systematically higher economic value than non-AI patents across all subfields, although the average level of patent value is lower than in the US sample. The relative AI premium is evident in all subfields. The largest differences appear in NLP, knowledge processing, and vision, suggesting a particularly strong market valuation of AI innovations in these subareas.

Both panels in the figure document a consistent cross-country pattern: AI patents are associated with higher market-based economic value than non-AI patents in both countries. At the same time, the AI valuation premium varies across technology subfields and across countries, with larger premia observed in data- and software-intensive AI technologies and smaller premia in more hardware-oriented areas. Importantly, the positive and systematic market valuation of Chinese AI patents is difficult to reconcile with a purely subsidy-driven view of patenting and suggests that, at least for listed firms, AI patents capture economically meaningful technological assets.

\begin{center}
    [Insert Figure~\ref{fig:cn_us_patent_value} About Here]
\end{center}

\paragraph{The Value of Innovation by Public Entities.}
A growing concern in the economics of AI is the shift of basic research from open academic institutions to proprietary corporate laboratories, often described as the ``privatization of science'' \citep{arora2018decline, jurowetzki2021privatization}. Although recent studies document a declining role for US universities in AI research \citep{jurowetzki2021privatization,gofman2024artificial}, the contribution of public entities in China remains less clear. A common concern is that Chinese universities and state-owned enterprises (SOEs), motivated in part by subsidy programs and administrative targets, may produce patents of limited economic relevance \citep{dang2015patent}.

We assess this concern by examining patterns of knowledge flows across sectors. If profit-oriented private firms draw on patents developed by SOEs and universities, these patents are likely to contain useful technological knowledge. Therefore, we study the behavior of citations using the notion of \textit{revealed technological preference}, which interprets citations by private firms as evidence of perceived technical value \citep{trajtenberg1990penny,alcacer2006patent,bloom2013identifying,jiao2021link}.

Table \ref{tab:cn_us_citation_assignee_type} reports the propensities for cross-sector citation. The results do not support the view that state-sector patents are largely ``junk''. Panel B shows that private enterprises cite patents held by SOEs and research institutions with a combined propensity of 0.533 ($0.212 + 0.321$), exceeding their propensity to cite patents from other private firms (0.467). Put differently, Chinese private firms are \textit{more} likely to build on inventions originating in the state sector than on those developed by private peers. This pattern is even stronger in AI-related technologies, where citations from private firms to SOEs ($0.321$) are more frequent than in non-AI fields ($0.287$), suggesting that state actors play an important upstream role in the development of economically relevant AI knowledge. 

Comparing institutional citation patterns across countries reveals two different patterns in knowledge flows between academia and industry. First, research institutions in the US are characterized by strong within-sector citation. As shown in Panel A, US institutions show a very high self-citation propensity (around 0.90) for both AI and non-AI patents, suggesting that academic research predominantly builds on prior academic work, with limited direct citation of industrial inventions. In comparison, Panel B shows that Chinese research institutions cite the corporate sector more frequently: for AI patents, citations to POEs and SOEs are 0.17 and 0.23, respectively. This pattern points to a more active cross-sector knowledge exchange in China \citep{etzkowitz2000dynamics}.

Second, despite the strong within-sector citation patterns observed for academia, firms in \textit{both} countries rely heavily on knowledge originating outside the private sector. US companies cite research institutions with a propensity of 0.475, while Chinese POEs cite state-sector patents with a combined propensity of 0.533 (=0.212+0.321). These patterns suggest that non-market institutions (universities and SOEs) provide important upstream knowledge that is subsequently drawn upon by private firms. More broadly, the evidence suggests that public-to-private knowledge flows play a central role in AI innovation in both economies, and provides evidence contrary to the view that non-market patents are primarily administrative in nature.

\subsection{Cross-Border AI Knowledge Dependency}
\label{sec:dependency}

Another important question is whether the US and China are becoming less connected in their technological development. \cite{Han2024} document a “hump-shaped” pattern in cross-border technological integration: links between the two countries strengthened in the early 2000s but have weakened in recent years. However, their analysis spans a wide range of technologies, including many mature industries where rising trade barriers and policy tensions may directly affect economic ties.

Instead, we focus on artificial intelligence, where the US and China compete directly in developing new technologies. We use this setting to study whether competition reduces cross-border learning or whether firms continue to build on each other’s innovations, using the learning-intensity framework of \cite{Han2024}.

We define a \textit{Relative Citation Propensity} index to measure relative knowledge sourcing. This index captures the degree to which patents in one country build upon the prior patents from the other country, relative to domestic patents, while accounting for the rapid expansion of the overall patent pool. Let $g \in \{\text{AI}, \text{Non-AI}\}$ denote the patent type. The propensity for type-$g$ Chinese patents to cite US patents relative to domestic patents is defined as:

\begin{equation}
p_{c,u,t}^g
=
\frac{n_{c,u,t}^g / x_{u,t}}{n_{c,c,t}^g / x_{c,t}},
\label{eq:citation_propensity_cu}
\end{equation}
where $n_{c,u,t}^g$ (respectively, $n_{c,c,t}^g$) denotes the count of citations made by Chinese patents of type $g$ to US (respectively, Chinese) patents of both types in year $t$, and $x_{u,t}$ (respectively, $x_{c,t}$) represents the cumulative number of US (respectively, Chinese) patents granted before year \textit{t}, which are eligible to be cited by Chinese patents in year \textit{t}.
Similarly, the propensity for type-$g$ US patents to cite Chinese patents relative to domestic patents is defined as:
\begin{equation}
p_{u,c,t}^g=\frac{n_{u,c,t}^g / x_{c,t}}{n_{u,u,t}^g / x_{u,t}},
\label{eq:citation_propensity_uc}
\end{equation}

Hence, a value of $p=1$ indicates ``technological neutrality'': Chinese patents cite US patents at the same rate as domestic patents, after accounting for the size of the available patent pool. Values above (or below) one indicate that Chinese patents systematically rely more on foreign (or domestic, respectively) technologies. Importantly, by normalizing the citation counts by cumulative patent stocks ($x_{u,t}$ and $x_{c,t}$), the measure separates changes in learning intensity from mechanical scale effects, ensuring that increases in citations driven solely by the rapid growth of China's patent activity are not misinterpreted as a deeper cross-border integration.

\begin{figure}[htbp]
\centering
\caption{Citation Propensity of AI and Non-AI Patents between the US and China}
\label{fig:cn_us_citation_propensity}
\caption*{
This figure plots the citation propensity defined in equations (\ref{eq:citation_propensity_cu}) and (\ref{eq:citation_propensity_uc}), following \cite{Han2024}, for cross-border citations between the US and China. The "any" sample includes patents that fall into at least one of the seven AI subcategories. For each AI subfield, we compute citation propensity separately for patents in that subfield and for non-AI patents, measuring their citations to all patents in the other country. Higher values indicate stronger technological reliance on foreign knowledge. For China, data from 2004-2006 are excluded due to small sample sizes.

}
\includegraphics[width=0.87\textwidth]{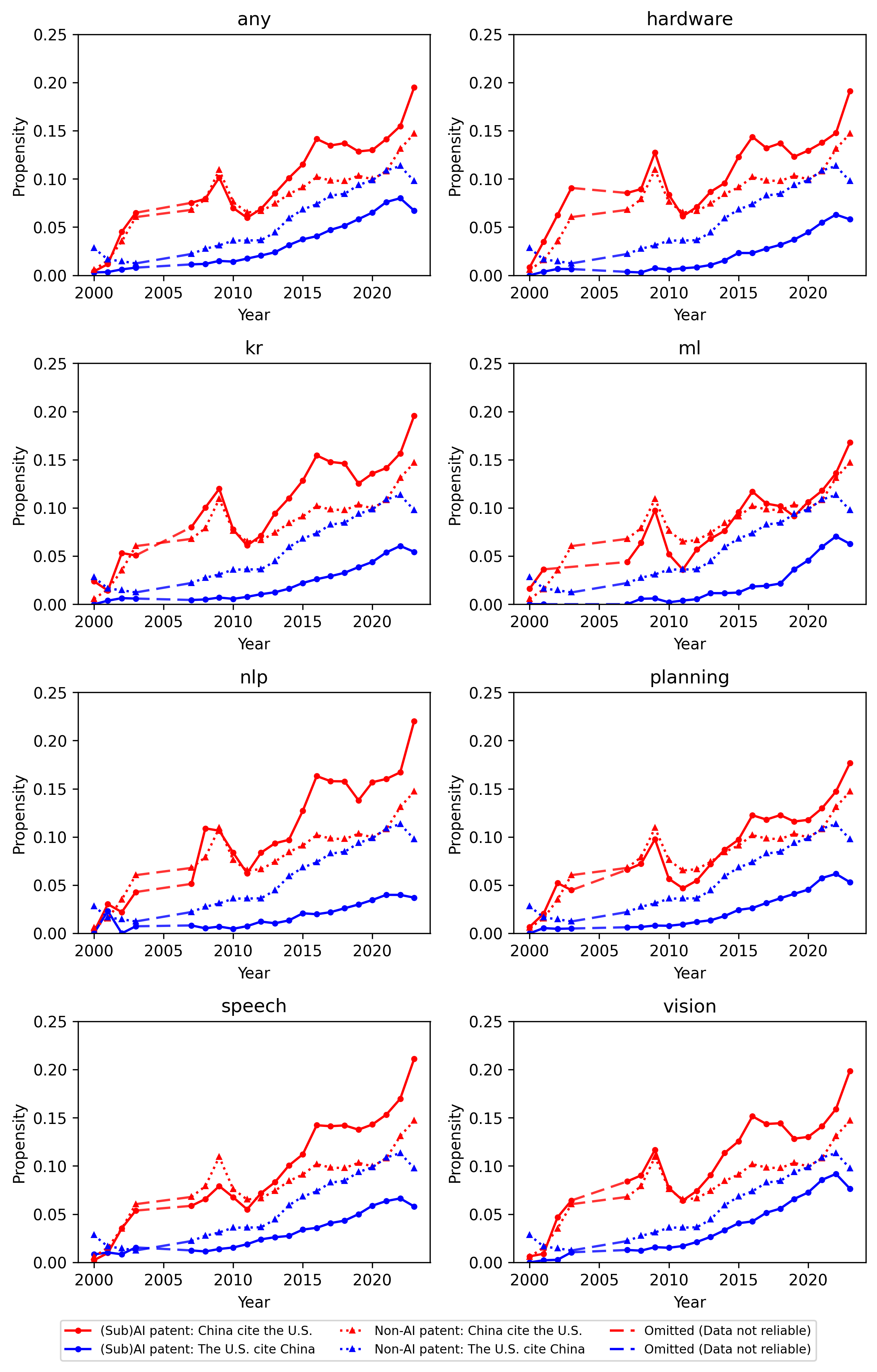}
\end{figure}

\begin{center}
    [Insert Figure~\ref{fig:cn_us_citation_propensity} About Here]
\end{center}

Figure \ref{fig:cn_us_citation_propensity} plots these propensity trends for AI versus Non-AI technologies. The figures reveal three facts about cross-border knowledge dependence between the two countries: First, the figures show a clear rise in cross-border citation intensity over time, for both countries and in both AI and non-AI domains. Despite considerable volatility in the early years—likely driven by small patent counts—the post-2005 patterns are steadily upward, suggesting a gradual increase in technological integration between the US and China. 

Second, even as the number of cross-border citations overall increases, the pattern in AI is uneven. Chinese AI patents cite US patents more intensively than Chinese non-AI patents do, indicating that reliance on US knowledge is particularly strong in AI. In contrast, US AI patents cite Chinese patents less intensively than US non-AI patents. 

Third, most subfields show a similar upward trend, but the imbalance between the two countries varies between domains. The gap relative to the non-AI benchmark is especially pronounced in algorithmic and data-intensive areas, whereas hardware-related fields appear more balanced. The overall AI pattern therefore seems to be driven primarily by a stronger directional dependence in certain domains.

\section{Conclusion}\label{sec:conclusion}

This paper develops a new high-resolution measure of AI patents by fine-tuning a domain-specific large language model on patent texts. A comprehensive set of validation exercises shows that the classifier achieves high precision and performs consistently across countries and subfields of artificial intelligence. Relative to existing USPTO classifications, our approach identifies a broader and more economically meaningful set of AI inventions, particularly in applied and hybrid domains that are often missed by rule-based or administrative definitions. 

Applying this measure to study AI innovations in the US and China, we uncover several robust patterns in the two countries. The US and China follow similar technological paths across AI domains, with China narrowing an early US lead over time. However, beneath this convergence, the structure and diffusion of AI innovation differ significantly. In the US, AI patenting is highly concentrated among a small set of large private firms and remains geographically concentrated in early hubs. In China, AI innovation spreads more rapidly across cities and is distributed across private firms, universities, and state-owned enterprises. Market-based evidence shows that AI patents command a clear valuation premium relative to non-AI patents in both economies, including those produced by non-market actors in China. Finally, citation patterns show continued cross-border learning rather than technological separation, with Chinese AI inventors relying heavily on US frontier technologies and more limited knowledge flows in the opposite direction.

\clearpage

\setlength{\bibsep}{0pt plus 0.3ex}

\bibliographystyle{apalike}
\bibliography{reference}
\restoregeometry

\newgeometry{left=1.in, right=1.in, top=1.in, bottom=1.in}
\newpage


\newpage


\newpage

\newpage

\newpage


\newpage

\newpage


\newpage


\newpage


\begin{figure}[htbp]
    \centering
    \caption{Geographic Distribution of AI Patents}
    \label{fig:cn_us_spatial_dynamic}
    \vspace{2mm}
    \caption*{This figure shows the geographic distribution of AI patents in the US and China across four time periods. Panel (a) plots the spatial locations of assignees for US AI patents, and Panel (b) presents the corresponding distribution for Chinese AI patents. Each map displays a kernel-based relative density surface, where larger markers indicate higher local concentrations of AI patenting activity. The densities are normalized to make intensity levels comparable across periods within each country.}
    \vspace{2mm}
    \makebox[\textwidth][c]{
    \begin{subfigure}[t]{1.2\textwidth}
        \centering
        \caption{US AI Patents}
        \includegraphics[width=\linewidth]
        {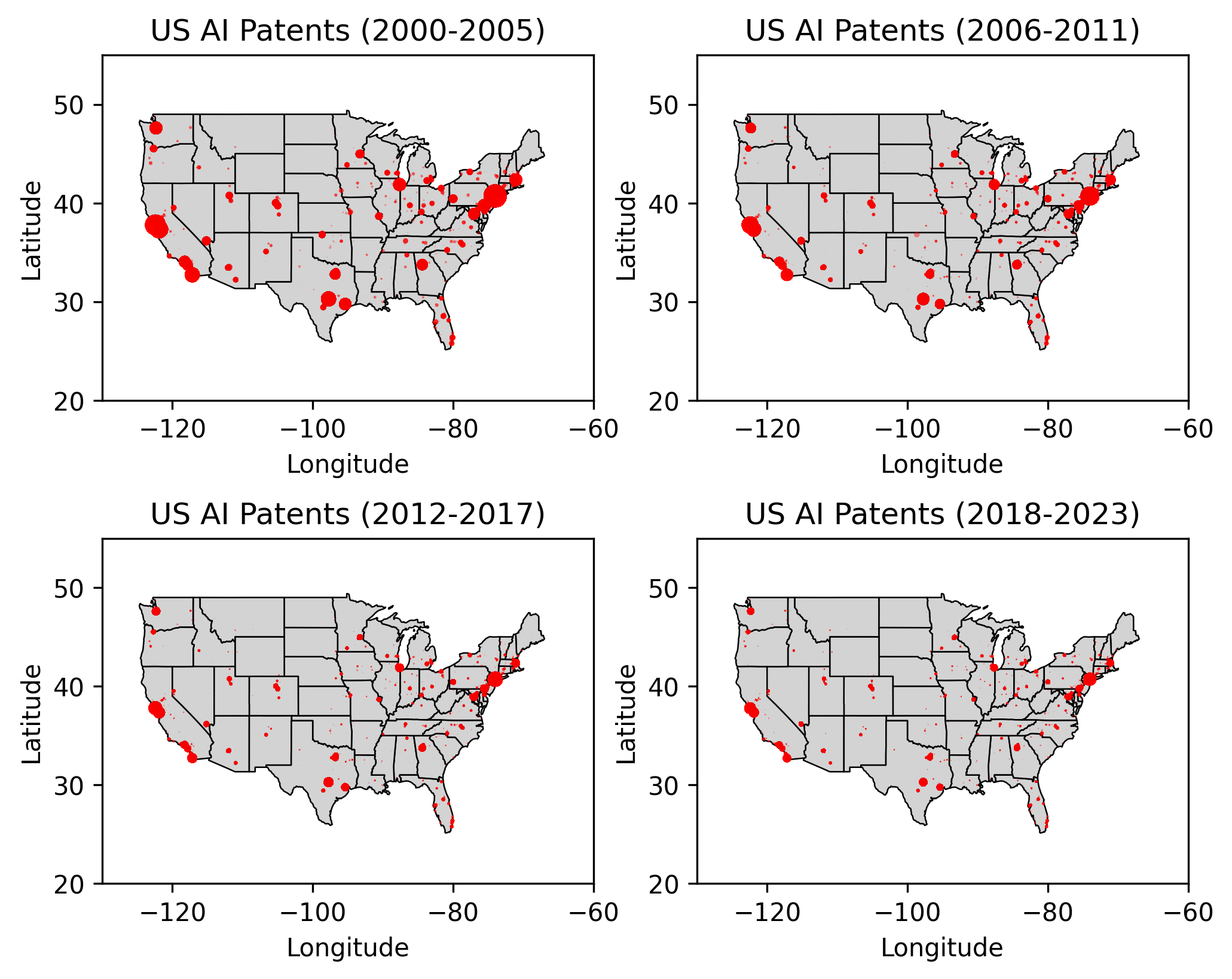}
    \end{subfigure}}
\end{figure}

\clearpage

\begin{figure}[htbp]
    \centering
    \ContinuedFloat  
    \vspace{2mm}
    \makebox[\textwidth][c]{
        \begin{subfigure}[t]{1.2\linewidth}
            \centering
            \caption{Chinese AI Patents}
            \includegraphics[width=\linewidth]{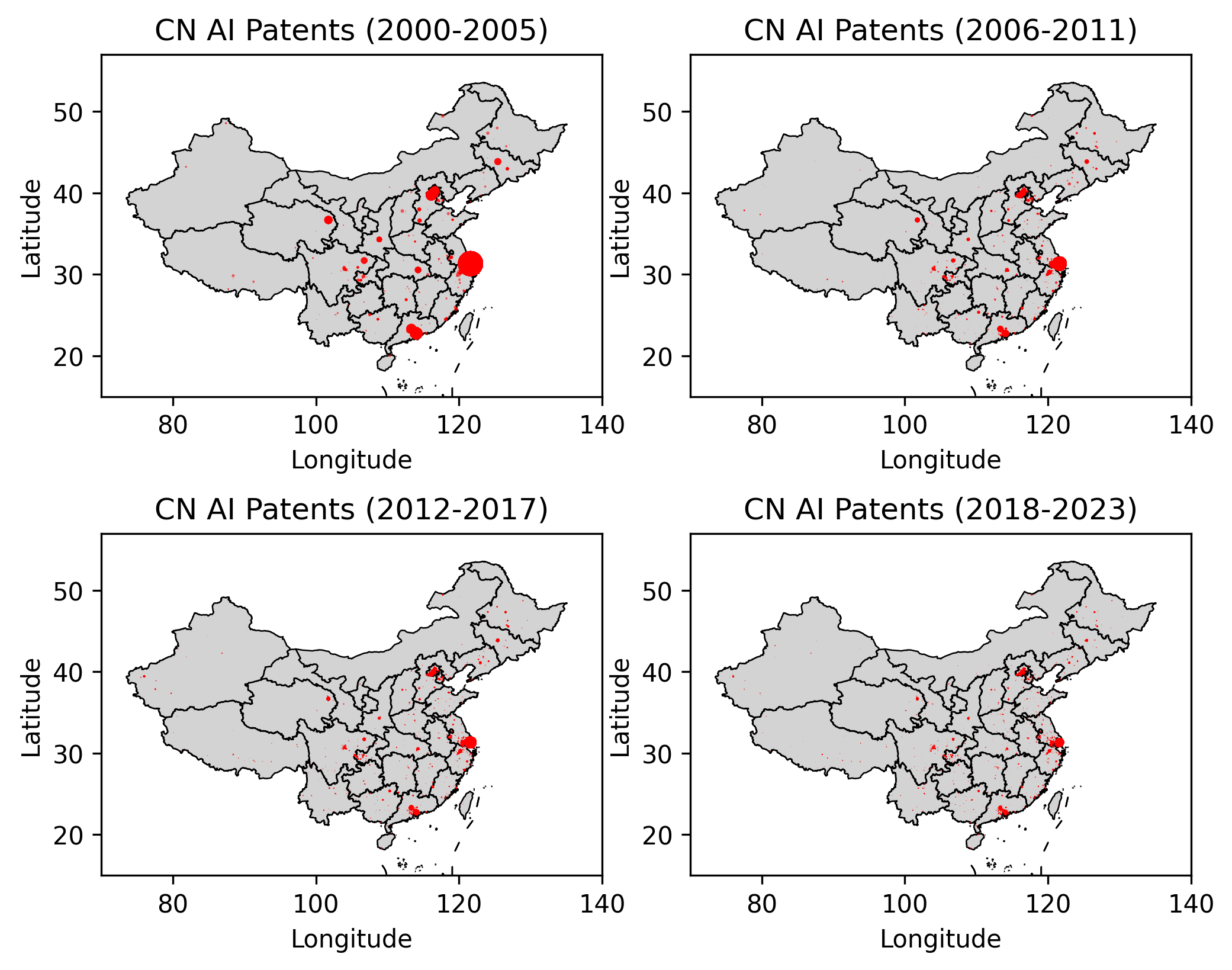}
        \end{subfigure}
    }
\end{figure}
\newpage


\newpage

\newgeometry{left=1.in, right=1.in, top=0.3in, bottom=1.in}
\restoregeometry
\clearpage


\clearpage

\newgeometry{top=0.3in, bottom=1.in}
\restoregeometry
\newpage

\begin{table}[htbp]
\centering
 \caption{Model Performance of Fine-tuned LLM Classifiers and Sample Composition}
\label{tab:train_and_test_accuracy}
\caption*{
This table provides a comprehensive overview of the fine-tuned classifier's performance and the underlying sample composition across AI patent subcategories. Panel A details the total number of samples, including positive ("seeds") and negative ("anti-seeds") instances, for both the training and test sets within each AI subcategory. Panel B presents key evaluation metrics for each subcategory, including \textit{Precision} (proportion of true positives among all positive predictions), \textit{Recall} (proportion of true positives among all actual positive samples), \textit{Accuracy} (overall proportion of correct predictions), and \textit{F1 score} (the harmonic mean of precision and recall). Panel C demonstrates a significant improvement in test performance by comparing our model's metrics against the initial LSTM model developed in \cite{giczy2022identifying}. The subcategory of \textit{evo}, due to an extremely limited number of labeled examples, shows weak predictive performance and is therefore excluded from all subsequent empirical analyses. ``NA" indicates metrics that are not calculated due to insufficient sample size.
}
\renewcommand{\arraystretch}{1.5}
\begin{subtable}[t]{\textwidth}
    \centering
    \caption{Panel A: Description of Training and Test Data}
    \makebox[\textwidth][c]{
    \begin{tabular}{p{3cm}*{8}{>{\centering\arraybackslash}p{1.5cm}}}
        \toprule
         & evo & hardware & kr & ml & nlp & planning & speech & vision \\
        \midrule
        Train Number   & 128  & 3,104   & 824 & 1,424 & 996  & 1,564    & 804    & 1,016   \\
        Train Positive  & 26   & 621     & 165 & 285   & 199  & 313      & 161    & 203    \\
        Train Negative  & 102  & 2,483   & 659 & 1,139 & 797  & 1,251    & 643    & 813    \\
        Test Number    & 32   & 776     & 206 & 356   & 249  & 391      & 201    & 254    \\
        Test Positive   & 6    & 155     & 41  & 71    & 50   & 78       & 40     & 51     \\
        Test Negative   & 26   & 621     & 165 & 285   & 199  & 313      & 161    & 203    \\
        \bottomrule
    \end{tabular}
    }
\end{subtable}

\vspace{0.5cm}

\begin{subtable}[t]{\textwidth}
    \centering
    \caption{Panel B: Model's Performance on Test Data}
    \makebox[\textwidth][c]{
    \begin{tabular}{p{3cm}*{8}{>{\centering\arraybackslash}p{1.5cm}}}
        \toprule
         & evo & hardware & kr & ml & nlp & planning & speech & vision \\
        \midrule
        Precision 
        & NA    & 0.963 & 0.975 & 1.000 & 1.000 & 0.974 & 0.976 &   0.911  \\
        Recall 
        & 0.000 & 0.839 & 0.951 & 0.958 & 0.940 & 0.949 & 1.000 & 1.000  \\
        Accuracy  
        & 0.813 & 0.961 & 0.985 & 0.992 & 0.988 & 0.985 & 0.995 & 0.980  \\
        F1 score  
        &   NA  & 0.897 & 0.963 & 0.978 & 0.969 & 0.961 & 0.988 & 0.953  \\
        \bottomrule
    \end{tabular}
    }
\end{subtable}

\vspace{0.5cm}

\begin{subtable}[t]{\textwidth}
    \centering
    \caption{Panel C: Overall Performance of Our Model v.s. LSTM Model by USPTO}
    \makebox[\textwidth][c]{
    \begin{tabular}{p{3cm}*{4}{>{\centering\arraybackslash}p{3cm}}}
        \toprule
         & Precision & Recall & Accuracy & F1 score \\
        \midrule
        Our Model  & 0.9698 & 0.9126 & 0.9769 & 0.9403 \\
        LSTM Model & 0.4054 & 0.3750 & 0.8723 & 0.3896 \\
        \bottomrule
    \end{tabular}
    }
\end{subtable}
\end{table}
\newpage


\begin{table}[htbp]
\centering
\caption{Top Assignees with Most AI Patents}
\label{tab:cn_us_assignee}
\caption*{
This table reports the top ten assignees in the US (Panel A) and China (Panel B) for each AI subfield, based on the number of \textit{FGYZ}-identified AI patents. Rankings are computed separately within each subfield.
}
\small
\renewcommand{\arraystretch}{1.5}

\begin{subtable}[t]{\textwidth}
\centering
\caption{Panel A: US}
\makebox[\textwidth][c]{
\begin{tabular}{*{8}{>{\centering\arraybackslash}p{2cm}}}
\toprule
 any & hardware & kr & ml & nlp & planning & speech & vision \\
\midrule
 IBM & IBM & IBM & IBM & IBM & IBM & IBM & IBM \\
 
Microsoft & Microsoft & Microsoft & Microsoft & Microsoft & Microsoft & Microsoft & Canon \\

Canon & Google & Google & Google & Google & Google & Google & Microsoft \\
 
Google & Intel & SAP & Samsung & SAP & Amazon & Samsung & Sony \\

Samsung & Amazon & Amazon & Amazon & Oracle & SAP & Sony & Google \\

Sony & SAP & Oracle & Intel & Amazon & Oracle & Amazon & Samsung \\

Amazon & Oracle & Samsung & Adobe & Facebook & AT\&T & AT\&T & Fujitsu \\
Intel & Samsung & Fujitsu & Capital One & Yahoo & Samsung & Apple & Toshiba \\

Fujitsu & Fujitsu & Facebook & NEC & AT\&T & Sony & Canon & Ricoh \\

Toshiba & NEC & HP & Fujitsu & Fujitsu & Intel & Fujitsu & Xerox \\
\bottomrule
\end{tabular}
}
\end{subtable}

\vspace{0.5cm}

\begin{subtable}[t]{\textwidth}
\centering
\caption{Panel B: China}
\makebox[\textwidth][c]{
\begin{tabular}{*{8}{>{\centering\arraybackslash}p{2cm}}}
\toprule
any & hardware & kr & ml & nlp & planning & speech & vision \\
\midrule
Tencent & Baidu & Tencent & Baidu & Baidu & Tencent & Tencent & Tencent \\

Baidu & Tencent & Baidu & Tencent & Tencent & Baidu & Baidu & Baidu \\

Huawei & UESTC & Alibaba & Zhejiang U & Microsoft & State Grid & Huawei & Xidian U \\

State Grid & Zhejiang U & Huawei & UESTC & Ping An & Huawei & Ping An & Huawei \\

Zhejiang U & Xidian U & Ping An & Xidian U & Alibaba & Zhejiang U & Samsung & UESTC \\

Tsinghua U & Tsinghua U & State Grid & Tsinghua U & IBM & BUAA & Alibaba & Zhejiang U \\

UESTC & BUAA & BUAA & State Grid & Google & Ping An & Microsoft & Tsinghua U \\

BUAA & State Grid & Zhejiang U & Ping An & Huawei & Tsinghua U & ZTE & BUAA \\

Xidian U & Huawei & Microsoft & BUAA & Adv. New & Alibaba & Tsinghua U & Ping An \\

Alibaba & Ping An & UESTC & SCUT & Tsinghua U & UESTC & Zhejiang U & Alibaba \\

\bottomrule
\end{tabular}
}
\end{subtable}
\end{table}
\newpage


\begin{table}[htbp]
    \centering
    \caption{Citation Propensity of AI and NonAI Patents between Different Assignee Types}
    \label{tab:cn_us_citation_assignee_type}
    \caption*{
    This table reports cross-type citation propensities for US and Chinese patents, comparing research institutions and enterprises in the US, and research institutions, private-owned enterprises (POEs), and state-owned enterprises (SOEs) in China. For each country and each patent category (AI and non-AI), rows indicate the type of the citing patent $i$ and columns indicate the type of the cited patent $j$. We define the citation propensity from patents of type $i$ to patents of type $j$ as the ratio of the share of citations from $i$ to $j$ to the share of patents belonging to type $j$, which is then normalized so that $\sum_j p_{i,j} = 1$ for each $i$. A higher value of $p_{i,j}$ indicates that patents of type $i$ cite patents of type $j$ more frequently than would be expected based on $j$'s overall patent share, whereas a lower value indicates relative under-citation.
    In particular, stronger diagonal elements reflect within-type knowledge concentration, while larger off-diagonal elements indicate cross-organizational knowledge flows and technological dependence across institutional boundaries.
    }
    \renewcommand{\arraystretch}{1.5} 
    \begin{subtable}[t]{\textwidth}
    \centering
    \caption{US Patents}
    \makebox[\textwidth][c]{
        \begin{tabular}{p{3cm}*{4}{>{\centering\arraybackslash}p{3cm}}}
            \toprule
             \multicolumn{1}{c}{} & \multicolumn{4}{c}{\textit{Citation Propensity}} \\
             \cmidrule(lr){2-5}
             &  \multicolumn{2}{c}{\textit{AI Patents}} & \multicolumn{2}{c}{\textit{Non-AI Patents}} \\
             \cmidrule(lr){2-3} 
             \cmidrule(lr){4-5} 
             & \textit{Institution} & \textit{Enterprise}  & \textit{Institution} & \textit{Enterprise}  \\
            \midrule
            \textit{Institution} & 0.898 & 0.102 & 0.904 & 0.096 \\
            \textit{Enterprise} & 0.475 & 0.525 & 0.458 & 0.542\\
            \bottomrule
        \end{tabular}
        }
    \end{subtable}
    
    \vspace{5mm}
    
    \begin{subtable}[t]{\textwidth}
    \centering
    \caption{Chinese Patents}
    \makebox[\textwidth][c]{
        \begin{tabular}{p{2cm}*{6}{>{\centering\arraybackslash}p{2cm}}}
            \toprule
             \multicolumn{1}{c}{} & \multicolumn{6}{c}{\textit{Citation Propensity}} \\
             \cmidrule(lr){2-7}
             &  \multicolumn{3}{c}{\textit{AI Patents}} & \multicolumn{3}{c}{\textit{Non-AI Patents}} \\
             \cmidrule(lr){2-4} 
             \cmidrule(lr){5-7} 
             & \textit{Institution} & \textit{POE} & \textit{SOE} & \textit{Institution} & \textit{POE} & \textit{SOE} \\
            \midrule
            \textit{Institution} & 0.596 & 0.171 & 0.233 & 0.639 & 0.157 & 0.204\\
            \textit{POE} & 0.212 & 0.467 & 0.321 & 0.207 & 0.506 & 0.287 \\
            \textit{SOE} & 0.229 & 0.274 & 0.497 & 0.209 & 0.215 & 0.576\\
            \bottomrule
        \end{tabular}
        }
    \end{subtable}
\end{table}
\newpage
\clearpage
\setcounter{table}{0}
\renewcommand{\thetable}{A\arabic{table}}
\setcounter{figure}{0}
\renewcommand{\thefigure}{A\arabic{figure}}
\setcounter{page}{1}
\renewcommand{\thepage}{A\arabic{page}}
\newgeometry{left=1.2cm, right=1.2cm, top=2cm, bottom=2cm}
\appendix
\section{Online Appendix A}
\subsection{Fine-Tuning Transformer Models for Patent Classification}\label{sec:technical-detail}

To classify patents into AI-related subfields, we fine-tune a domain-specific transformer model using a supervised deep learning framework implemented in PyTorch Lightning. Our approach builds on \textit{PatentSBERTa}, a Sentence-BERT (SBERT) architecture specifically adapted to the semantic structure of patent texts. Originally developed by \citet{bekamiri2021patentsberta}, \textit{PatentSBERTa} was pretrained on large-scale patent corpora to generate semantically meaningful embeddings of patent claims and abstracts, and subsequently fine-tuned to predict Cooperative Patent Classification (CPC) labels. This pretraining strategy ensures that the model captures the technical vocabulary, structure, and style unique to patent documents, outperforming general-purpose language models such as BERT and RoBERTa on intellectual property and innovation tasks.

We adapt this pretrained model and use the PatentSBERTa as our base model for AI patent classification by attaching a task-specific classification head. The architecture consists of two components: (i) the unfrozen transformer encoder from \textit{PatentSBERTa}, which generates contextualized vector representations of patent abstracts, and (ii) a feedforward classification module with fully connected layers, ReLU activations, and dropout regularization. The final layer outputs a softmax probability over binary class labels, indicating whether a given patent belongs to a particular AI subdomain (e.g., machine learning, computer vision, natural language processing).

\paragraph{Architecture and Training Objective.}
Let ${x}$ denote a tokenized patent document (abstract and claims). We model the patent encoder with a pretrained transformer (PatentSBERTa model with 109 M parameters), denoted
\[
g({x};\,\theta)\;:\; {x}\mapsto h\in\mathbb{R}^{d},
\]
where $\theta$ are encoder parameters and $h$ is the pooled representation for the patent document.

On top of $h$, we place a task–specific classifier $f(\cdot;\gamma)$ implemented as a \emph{width–halving} multi–layer perceptron (MLP) with ReLU and dropout, followed by a final linear map to $K$ classes. Write the input width as $d_0=d$. For a user–specified depth $L\in\{0,1,2,\dots\}$, define layer widths recursively by
\[
d_\ell \;=\; \left\lfloor \frac{d_{\ell-1}}{2}\right\rfloor\!, \qquad \ell=1,\dots,L.
\]
Let $a^{(0)}=h$. For $\ell=1,\dots,L$, compute
\[
u^{(\ell)} \;=\; W^{(\ell)} a^{(\ell-1)} + b^{(\ell)} \in \mathbb{R}^{d_\ell}, 
\qquad 
\tilde{a}^{(\ell)} \;=\; \operatorname{ReLU}\!\big(u^{(\ell)}\big),
\qquad 
a^{(\ell)} \;=\; \operatorname{Dropout}_{p}\!\big(\tilde{a}^{(\ell)}\big),
\]
where $W^{(\ell)}\in\mathbb{R}^{d_\ell\times d_{\ell-1}}$, $b^{(\ell)}\in\mathbb{R}^{d_\ell}$, and $p\in[0,1)$ is the dropout rate. The classifier’s final linear layer maps $a^{(L)}$ to logits
\[
z \;=\; W^{(L+1)} a^{(L)} + b^{(L+1)} \in \mathbb{R}^{K},
\]
with $W^{(L+1)}\in\mathbb{R}^{K\times d_L}$ and $b^{(L+1)}\in\mathbb{R}^{K}$. Collect all classifier parameters as $\gamma=\{W^{(\ell)},b^{(\ell)}\}_{\ell=1}^{L+1}$ where \(\gamma\) is roughly contains 1.5k parameters. The full model thus composes encoder and head:
\[
z \;=\; f\!\big(g(x,m;\theta);\gamma\big).
\]

\paragraph{Binary and Multi–Class Outputs.}
For binary AI vs. non–AI classification ($K=2$), the softmax probabilities are
\[
\hat{\pi}_k \;=\; \frac{\exp(z_k)}{\sum_{j=1}^{2}\exp(z_j)} , \quad k\in\{1,2\}.
\]
Equivalently, with a single logit (log–odds) $z\in\mathbb{R}$ one may use $\sigma(z)=1/(1+e^{-z})$ for the AI class. For multi–label extensions with $K$ AI subdomains, apply elementwise sigmoid, $\hat{p}_k=\sigma(z_k)$.

\paragraph{Loss and Optimization.}
Given a minibatch $\{(x_i,y_i)\}_{i=1}^B$ with one–hot labels $y_i\in\{0,1\}^K$ (binary case: $K=2$), we minimize the cross–entropy loss
\[
\mathcal{L}(\theta,\gamma)
\;=\;
-\frac{1}{B}\sum_{i=1}^{B}\sum_{k=1}^{K} y_{ik}\,\log \hat{\pi}_{ik}
\;+\;
\lambda_\theta \|\theta\|_2^2
\;+\;
\lambda_\gamma \|\gamma\|_2^2,
\]
where $\hat{\pi}_{ik}$ are softmax probabilities from $z_i=f(g(x_i,m_i;\theta);\gamma)$, and $(\lambda_\theta,\lambda_\gamma)\ge 0$ are $L_2$ regularization weights. 
We train end–to–end with AdamW, learning rate $\eta$, and a linear warm–up over the first $\rho\in(0,1)$ fraction of steps. Early stopping selects the checkpoint with the lowest validation loss.

\paragraph{Decision Rule and Metrics.}
Predicted class $\hat{y}$ is the argmax of logits ($K\ge 2$). We report accuracy, precision, recall, and $F_1$ on held–out data.

\paragraph{Summary.}
Formally, the fine–tuned AI patent classifier is the composition
\[
\boxed{\quad z \;=\; f\!\big(g(x;\theta);\gamma\big)\,, \qquad \hat{\pi}=\operatorname{softmax}(z)\quad}
\]
where $g(\cdot;\theta)$ is the pretrained PatentSBERTa encoder and $f(\cdot;\gamma)$ is the width–halving ReLU–dropout MLP head described above. This mathematical specification matches our implementation: the encoder provides ($h$), which is passed through $L$ (we set \(L = 3\) at most and select it as a hyperparameter) halving linear layers with ReLU and dropout, and a final linear layer produces logits for cross–entropy training.

Each subdomain is framed as an independent binary classification problem. For this purpose, we compile labeled datasets from expert-annotated patent samples spanning nine major AI fields. These datasets are randomly split into training (80\%) and validation (20\%) sets using stratified sampling to preserve class balance.

Training proceeds for up to 20 epochs per subdomain. We optimize with AdamW, using a fixed learning rate of $2 \times 10^{-5}$ and a linear warm-up schedule over the first 20\% of steps to stabilize convergence. We employ a batch size of 32 and monitor validation accuracy in each epoch. To mitigate overfitting, we apply early stopping with a patience of three epochs, retaining the model checkpoint with the highest validation performance for evaluation.

Performance is assessed on held-out validation sets using accuracy, precision, recall, and F1-score, thereby capturing both correctness and robustness across heterogeneous subdomains that differ in linguistic style, technical content, and semantic granularity.

The choice of \textit{PatentSBERTa} reflects the growing consensus that domain-specific pretraining substantially improves downstream performance on technical and scientific corpora \citep{beltagy2019scibert}. Leveraging a model already adapted to patent language ensures that the learned representations align closely with the specialized vocabulary and structure of innovation disclosures—an essential feature for reliable classification in high-stakes policy and economic analysis.

\newpage
\subsection{AI Patent Definitions and Examples}
\label{tab:AIpatent_definition_example}
\cite{bekamiri2021patentsberta} categorizes AI-related patents into eight distinct types, and develops unique definitions for each category, delivering in-depth elaborations to delineate their technical meanings and application boundaries. We provide relevant illustrations here. Furthermore, a specific abstract example is furnished for each type of AI patent, which serves as a tangible presentation to facilitate understanding.
\begin{table}[htbp]
  \centering
  \small
  \begin{tabular}{p{2cm}p{5cm}p{11cm}}
    \toprule
    \textbf{Type} & \textbf{Illustration} & \textbf{Example(abstract)} \\
    \midrule
    Knowledge processing (kr) & The field of knowledge processing contains methods to represent facts about the world and to derive new facts (or knowledge) from a knowledge base. &  The performance of a given computer task is optimized by utilizing a plurality of intelligent agents suited to perform the computer task but having varied degrees of domain knowledge. Based upon an objective criteria that may be determined for a given situation, one of the plurality of intelligent agents may be selected and dispatched to perform the task, thereby optimizing the performance of the computer task for a wide variety of situations. (US6192354B1) \\ 
    \midrule
    Speech (speech) & Speech recognition includes methods to understand a sequence of words given an acoustic signal. It often integrates signal processing techniques to reduce interference from factors like background noise. &  A computer-implemented method and an apparatus are provided. The method includes obtaining, by a processor, a frequency spectrum of an audio signal data. The method further includes extracting, by the processor, periodic indications from the frequency spectrum. The method also includes inputting, by the processor, the periodic indications and components of the frequency spectrum into a neural network. The method additionally includes estimating, by the processor, sound identification information from the neural network. (US10062378B1)\\
    \midrule
    AI hardware (hardware) & The field of AI hardware includes physical hardware designed to implement artificial intelligence software. It may include logic circuitry, memory, video, processors, and solid-state technologies. & A number of consecutive samples in unit distance code (here Gray code) are effectively stacked and supplied to respective sum and threshold devices 20 corresponding to each bit position, to determine the bulk property or “generic result” of the consecutive samples. This filtered output is converted back to binary and analogue if required. The digital filter may be used in any of the above applications to “clean” data. (US6519577B1) \\
    \midrule
    Evolutionary computation (evo)  & Evolutionary computation contains a set of computational methods utilizing aspects of nature and, specifically, evolution. & A multi-objective optimization method. The method comprises a population of objects for each objective utilizing an optimization process and determining a measure for the variation in values of each gene within each population. A crossbreed between objects from different ones of the populations is performed, wherein a selection of gene values for the child object is weighted based on the measures for the variations in the values of each gene within the respective populations. (US9047569B2) \\
    \bottomrule
  \end{tabular}
\end{table}
\begin{table}[htbp]
  \centering
  \small
  \begin{tabular}{p{2cm}p{5cm}p{11cm}}
    \toprule
    \textbf{Type} & \textbf{Illustration} & \textbf{Example(abstract)} \\
    \midrule
    Natural language processing (nlp) & Natural language processing contains methods for understanding and using data encoded in human natural language. & A system, method and computer readable medium for providing translated web content is disclosed. The method on an information processing system includes retrieving a first content in a first language and parsing the first content into a plurality of translatable components. The method further includes generating a unique identifier for each of the plurality of translatable components of the first content and queuing the plurality of translatable components and corresponding unique identifiers for translation into a second language. The method further includes, for each of the plurality of translatable components, storing a translated component and an associated unique identifier corresponding to the translatable component, thereby storing a plurality of translated components and corresponding unique identifiers. (US7627479B2) \\
    \midrule
    Machine learning (ml) & The field of machine learning contains a broad class of computational learning models, including supervised learning classification models, neural networks, fuzzy logic, adaptive systems, probabilistic networks, regression and intelligent searching. & A neural network, which may be implemented either in hardware or software, is constructed of neurons or neuron circuits each having only one significant processing element in the form of an adder. Each neural network further includes circuits for applying a logarithmic function to its inputs and for applying an inverse-logarithmic function to the outputs of its neurons. The neural network utilizes a training algorithm which does not require repetitive training and which yields a global minimum to each given set of input vectors. (US5778153A) \\
    \midrule
    Computer vision  (vision) & The field of computer vision contains methods to extract and understand information from visual input, including images and videos. Areas of computer vision may include object recognition, manipulation (e.g., transformation, enhancement, or restoration), color processing, and conversion.  & The present invention relates to a method and system for detecting biologically relevant structures in a hierarchical fashion, beginning at a low-resolution and proceeding to higher levels of resolution. The present invention also provides probabilistic pairwise Markov models (PPMMs) to classify these relevant structures. The invention is directed to a novel classification approach which weighs the importance of these structures. The present invention also provides a fast, efficient computer-aided detection/diagnosis (CAD) system capable of rapidly processing medical images (i.e. high throughput). The computer-aided detection/diagnosis (CAD) system of the present invention allows for rapid analysis of medical images the improving the ability to effectively detect, diagnose, and treat certain diseases. (US8718340B2) \\
    \midrule
    Planning/ control (planning) & The field of planning and control contains methods to identify and execute plans to achieve specified goals. Key aspects of planning include representing actions and states of the world, reasoning about the effects of actions, and efficiently searching over potential plans.  & In an embodiment, a processor includes a fuzzy thermoelectric cooling (TEC) controller to: obtain a current TEC level associated with the processor; obtain a current fan power level associated with the processor; fuzzify the current TEC level to obtain a first fuzzy fan level; fuzzify the current fan power level to obtain a second fuzzy fan level; determine a new TEC power level based at least in part on the first fuzzy fan level, the second fuzzy fan level, and a plurality of fuzzy rules; and provide the new TEC power level to a TEC device associated with the processor, where the TEC device is to transfer heat from the processor to a heat sink. Other embodiments are described and claimed. (US9857809B2) \\
    \bottomrule
  \end{tabular}
\end{table}

\newpage

\subsection{AI Patent with Multiple Subcategories}
\label{fig:subcategory_percentage}
\begin{figure}[htbp]
    \centering
    \vspace{2mm}
    \caption*{ This figure presents the distribution of AI patents by the number of AI subfields to which they belong. We train eight classifiers, each corresponding to a specific AI subcategory, and apply them to patent texts to determine whether a patent belongs to each subcategory. Because these classifications are not mutually exclusive, a single patent may be assigned to multiple AI subfields. The histogram reports the percentage of AI patents that fall into one, two, three, or more subfields. The evolutionary computation category (\textit{evo}) is excluded, so the number of subfields ranges from 1 to 7. ``US-AIPD'' refers to classifications provided by the USPTO AI Patent Dataset, while ``US-FGYZ'' and ``China-FGYZ'' denote classifications generated by our FGYZ classifier for U.S. and Chinese patents, respectively.}
    \vspace{2mm}
    \includegraphics[width=\textwidth]{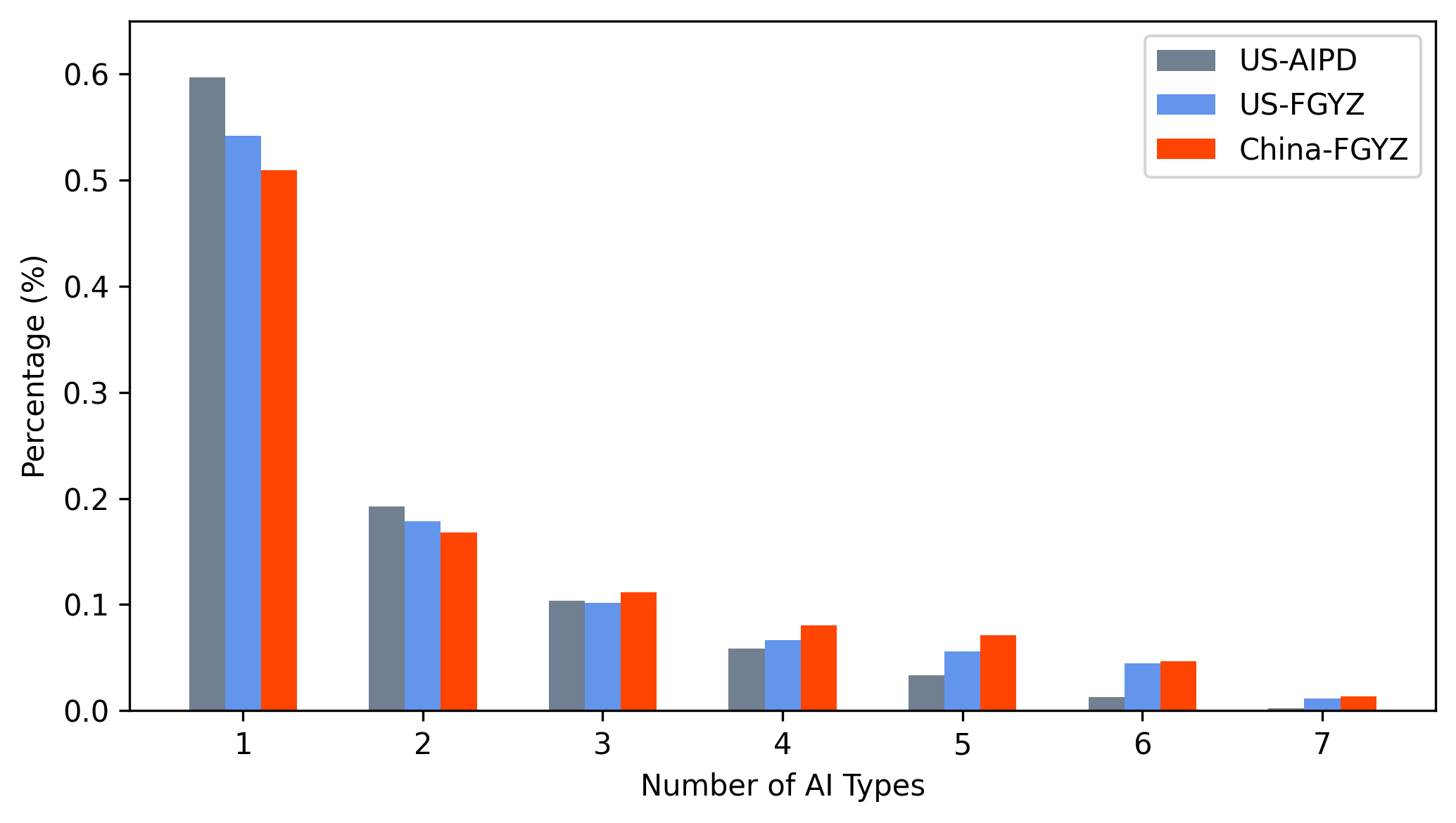}
\end{figure}

\end{document}